\documentclass[english, 11pt]{article}
\usepackage[T1]{fontenc}
\usepackage[utf8]{inputenc}
\usepackage{geometry}
\usepackage{float}
\usepackage{graphicx}
\usepackage{babel}
\usepackage{bm}
\usepackage{amsmath, amsthm}
\usepackage{amssymb} 
\usepackage{bm}
\usepackage{fixmath} 
\usepackage{subcaption}
\usepackage[cal=cm]{mathalfa}
\usepackage{algorithm2e}
\usepackage{caption}
\usepackage{subcaption}
\usepackage{xcolor}
\usepackage[round, sort&compress,comma,numbers]{natbib}
\usepackage{authblk}
\usepackage[normalem]{ulem}
\usepackage{dsfont}
\usepackage{cellspace}
\useunder{\uline}{\ul}{}
\RestyleAlgo{ruled}
\LinesNumbered
\usepackage{multirow}
\usepackage{tikz}
\usepackage{subcaption}
\usepackage{adjustbox}
\usepackage{calc}
\usetikzlibrary{positioning, shapes, calc, fit}

\usepackage{hyperref} 
\hypersetup{colorlinks=true,linkcolor=blue,filecolor=blue,citecolor=blue}
\newcommand{\indep}{\perp \!\!\! \perp}
\newcommand{\logit}{\text{logit}}

\makeatletter
\let\save@mathaccent\mathaccent
\newcommand*\if@single[3]{%
  \setbox0\hbox{${\mathaccent"0362{#1}}^H$}%
  \setbox2\hbox{${\mathaccent"0362{\kern0pt#1}}^H$}%
  \ifdim\ht0=\ht2 #3\else #2\fi
  }
\newcommand*\rel@kern[1]{\kern#1\dimexpr\macc@kerna}
\newcommand*\widebar[1]{\@ifnextchar^{{\wide@bar{#1}{0}}}{\wide@bar{#1}{1}}}
\newcommand*\wide@bar[2]{\if@single{#1}{\wide@bar@{#1}{#2}{1}}{\wide@bar@{#1}{#2}{2}}}
\newcommand*\wide@bar@[3]{%
  \begingroup
  \def\mathaccent##1##2{%
    \let\mathaccent\save@mathaccent
    \if#32 \let\macc@nucleus\first@char \fi
    \setbox\z@\hbox{$\macc@style{\macc@nucleus}_{}$}%
    \setbox\tw@\hbox{$\macc@style{\macc@nucleus}{}_{}$}%
    \dimen@\wd\tw@
    \advance\dimen@-\wd\z@
    \divide\dimen@ 3
    \@tempdima\wd\tw@
    \advance\@tempdima-\scriptspace
    \divide\@tempdima 10
    \advance\dimen@-\@tempdima
    \ifdim\dimen@>\z@ \dimen@0pt\fi
    \rel@kern{0.6}\kern-\dimen@
    \if#31
      \overline{\rel@kern{-0.6}\kern\dimen@\macc@nucleus\rel@kern{0.4}\kern\dimen@}%
      \advance\dimen@0.4\dimexpr\macc@kerna
      \let\final@kern#2%
      \ifdim\dimen@<\z@ \let\final@kern1\fi
      \if\final@kern1 \kern-\dimen@\fi
    \else
      \overline{\rel@kern{-0.6}\kern\dimen@#1}%
    \fi
  }%
  \macc@depth\@ne
  \let\math@bgroup\@empty \let\math@egroup\macc@set@skewchar
  \mathsurround\z@ \frozen@everymath{\mathgroup\macc@group\relax}%
  \macc@set@skewchar\relax
  \let\mathaccentV\macc@nested@a
  \if#31
    \macc@nested@a\relax111{#1}%
  \else
    \def\gobble@till@marker##1\endmarker{}%
    \futurelet\first@char\gobble@till@marker#1\endmarker
    \ifcat\noexpand\first@char A\else
      \def\first@char{}%
    \fi
    \macc@nested@a\relax111{\first@char}%
  \fi
  \endgroup
}
\makeatother

\begin{document}

\title{Adjusting for Selection Bias Due to Missing Eligibility Criteria in Emulated Target Trials}
\author[1]{Luke Benz}
\author[1]{Rajarshi Mukherjee}
\author[1,2,3]{Rui Wang}
\author[4]{David Arterburn}
\author[5]{Heidi Fischer}
\author[6]{Catherine Lee}
\author[7,8]{Susan M. Shortreed}
\author[1]{Sebastien Haneuse}
\affil[1]{Department of Biostatistics,
Harvard T.H. Chan School of Public Health, Boston, MA, USA}
\affil[2]{Department of Population Medicine, Harvard Pilgrim Health Care Institute, Boston, MA, USA}
\affil[3]{Department of Population Medicine, Harvard Medical School, Boston, MA, USA}
\affil[4]{Kaiser Permanente Washington Health Research Institute, Seattle, WA, USA}
\affil[5]{Department of Research \& Evaluation, Kaiser Permanente Southern California, Pasadena, CA, USA}
\affil[6]{Department of Epidemiology and Biostatistics, University of California San Francisco, San Francisco, CA, USA}
\affil[7]{Biostatistics Division, Kaiser Permanente Washington Health Research Institute, Seattle, WA, USA}
\affil[8]{Department of Biostatistics, University of Washington School of Public Health, Seattle, WA, USA}

\date{
    \today 
}

\maketitle

\begin{abstract}

\noindent 
Target trial emulation (TTE) is a popular framework for observational studies based on electronic health records (EHR). A key component of this framework is determining the patient population eligible for inclusion in both a target trial of interest and its observational emulation. Missingness in variables that define eligibility criteria, however, presents a major challenge towards determining the eligible population when emulating a target trial with an observational study. In practice, patients with incomplete data are almost always excluded from analysis despite the possibility of selection bias, which can arise when subjects with observed eligibility data are fundamentally different than excluded subjects. Despite this, to the best of our knowledge, very little work has been done to mitigate this concern. In this paper, we propose a novel conceptual framework to address selection bias in TTE studies, tailored towards time-to-event endpoints, and describe estimation and inferential procedures via inverse probability weighting (IPW). Under an EHR-based simulation infrastructure, developed to reflect the complexity of EHR data, we characterize common settings under which missing eligibility data poses the threat of selection bias and investigate the ability of the proposed methods to address it. Finally, using EHR databases from Kaiser Permanente, we demonstrate the use of our method to evaluate the effect of bariatric surgery on microvascular outcomes among a cohort of severely obese patients with Type II diabetes mellitus (T2DM).

\newpage{}
\end{abstract}




\section*{Introduction}
\label{sec:introduction}

Target trial emulation (TTE) has recently gained popularity as a framework for planing and conducting observational studies using electronic health records (EHR) and other routinely collected data. The premise of TTE is that most observational studies examining comparative effectiveness can be framed as attempts to emulate a (hypothetical) randomized trial, the so-called ``target trial'', that would answer the causal question of interest. This framework entails explicitly specifying the target trial through its protocol components (including eligibility criteria and the start of follow-up) and analyzing the observational data in a manner that adheres as closely as possible to the target trial specification. TTE offers a principled road map for thinking about and addressing challenges of observational studies, including confounding \cite{hernan2024}, immortal-time bias \citep{hernan2016immortal, agarwal2019immortal}, non-adherence, and loss to follow-up \cite{hernan2004structural}. The TTE framework has been extensively applied in the study of coronary heart disease and cardiovascular surgery \citep{hernan2008, danaei2013observational, danaei2018, massol2023levosimendan, matthews2021comparing, matthews2022benchmarking}, HIV \citep{cain2010, gran2010sequential, caniglia2019}, cancer \citep{emilsson2018, dickerman2019, petito2020, maringe2020reflection, kwee2023}, reproductive health \citep{yland2022, chiu2022, caniglia2023}, COVID-19 \citep{gupta2021, barda2021,  hoffman2022, dickerman2022, hajage2022extracorporeal, mcconeghy2024early}, and other areas \citep{clark2015, rossides2021, scola2023implementation}. 

In EHR data, patients may frequently meet the eligibility criteria for inclusion in a study population at many points in time, and determining where to begin follow-up time can be challenging, particularly for non-initiators of treatment. Emulating a target trial using observational data that chooses a single fixed time point and then follows eligible subjects who happened to initiate or not initiate treatment at that time can address this challenge \citep{hernan2016immortal}, but would exclude much of the available data (e.g. when  few people receive one of the of treatments of interest at any given time point the analyses cannot produce precise estimates for contrasts that involve that treatment). To address this, many works pursuing TTE have used a sequential variation of this framework when analyzing time-to-event outcomes \citep{hernan2008, cain2010, danaei2013observational, clark2015, emilsson2018, danaei2018, maringe2020reflection, gupta2021, hajage2022extracorporeal, massol2023levosimendan, caniglia2023}. Briefly, a series of target trials are conducted at regularly occurring intervals (e.g. monthly or annually) and pooled together in a much larger data set.

\begin{figure}[h]
    \centering
    \includegraphics[width=0.7\textwidth]{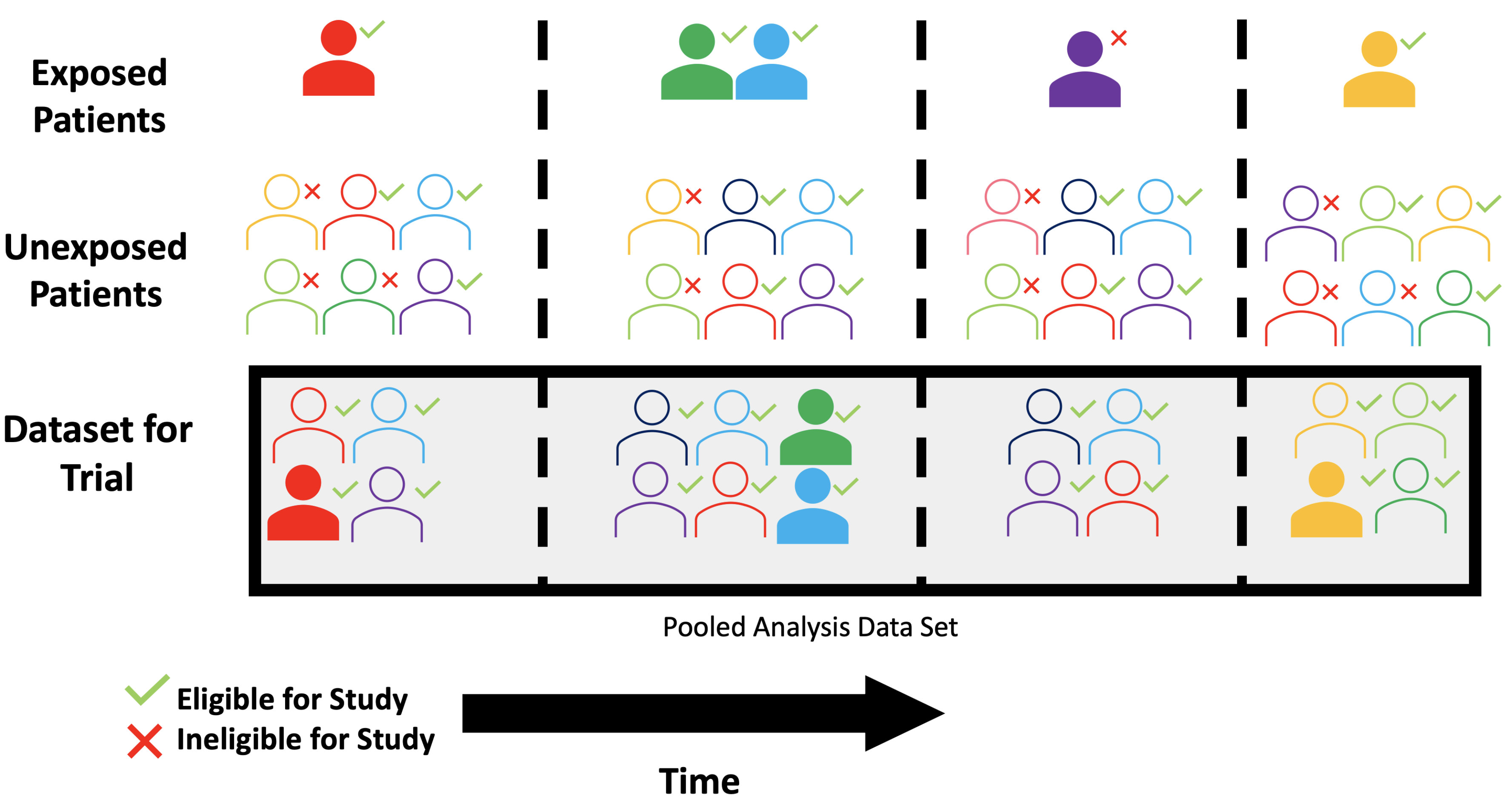}
    \caption{Overview of the sequential target trial emulation design. At regular time intervals, subjects are classified initiators of an exposure during that interval or non-initiators. Eligible subjects are ascertained to comprise an analysis data set for each target trial. While each trial could be analyzed individually, data are commonly pooled across trials into a single analysis dataset.}
    \label{fig:seq-tte}
\end{figure}

As illustrated in Figure \ref{fig:seq-tte}, determining which subjects meet eligibility criteria for inclusion in the study population is a critical step. Because EHR data are created for clinical and billing purposes, not all covariates relevant to the research question of interest are routinely collected in clinical care, and such information may not be collected consistently across time \citep{haneuse2016a}. Therefore, missing data in variables that define study eligibility criteria presents an important challenge. In practice, patients with incomplete eligibility data are frequently excluded from analysis \citep{danaei2013observational} despite the possibility of selection bias. Alternatively, eligibility status may be imputed via a patient's most recent covariate value(s) \citep{obrien2018microvascular}, even though such value(s) may not correctly reflect a patient's current eligibility status (Figure \ref{fig:ehr-durable}). 

Despite the popularity of the target trial framework, very few works have considered the problem  of selection bias due to missing eligibility criteria \citep{tompsett2023target}. Tompsett et al. proposed multiple imputation of eligibility-defining criteria and demonstrated its utility in addressing selection bias \cite{tompsett2023target}. In contrast to the work of Tompsett et al., which focuses on a single target trial and continuous outcome, we focus on time-to-event endpoints under a sequence of target trials and the added complexities that are introduced in this setting. In this work, we outline an inverse probability weighting (IPW) framework to address selection bias in TTE studies tailored towards time-to-event endpoints. Given that one of the first steps in many retrospective EHR-based observational studies is determining the study eligible population for analysis, any setting where eligibility is not readily ascertainable for all subjects might be susceptible to selection bias. Though the focus of this work is the sequential TTE design, the challenges and solutions explored in this work could apply more broadly, and as such, this work aims to fill an important gap in the literature. 

\section*{Bariatric Surgery}
\label{sec:bariatric_surgery}

\begin{figure}[ht]
    \centering
    \includegraphics[width=\textwidth]{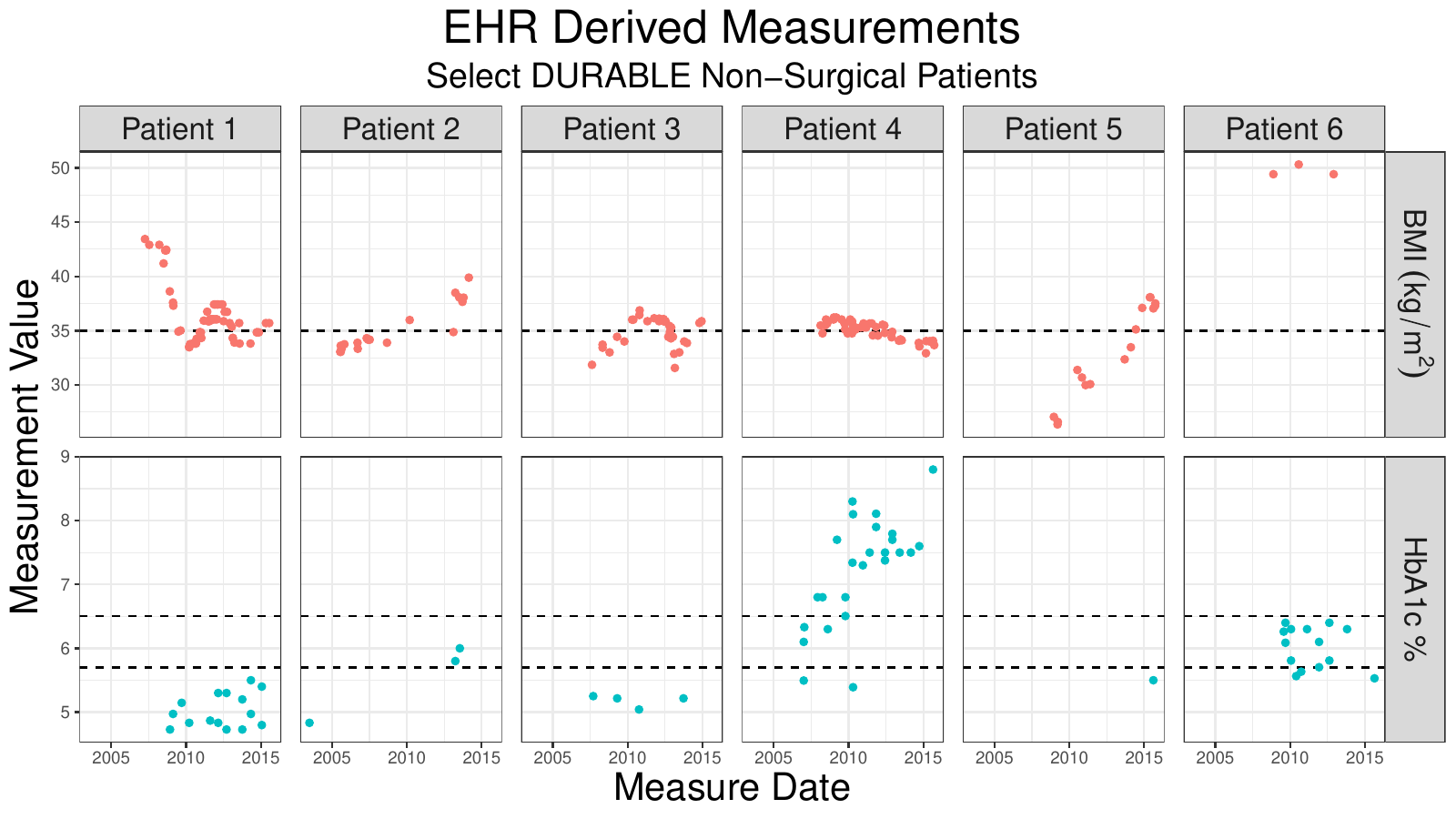}
    \caption{Hemoglobin A1c (HbA1c, blue dots) and body mass index (BMI, red dots) measurements over time for six non-surgical patients in the DURABLE study. Measurements are shown in relation to a clinically relevant  BMI threshold for consideration of bariatric surgery (35 kg/m$^2$), as well as A1c cutoffs defining prediabetes (5.7\%) and diabetes (6.5\%).}
    \label{fig:ehr-durable}
\end{figure}

Our work is motivated by a series of EHR-based observational studies to understand long-term outcomes following bariatric surgery \citep{arterburn2014bariatric, arterburn2020, coleman2016longterm, coleman2022bariatric, courcoulas2013weight, koffman2021investigating, koffman_evaluation2021,  obrien2018microvascular}. To illustrate the complexity of typical EHR data, and thus, potential challenges in determining eligibility when emulating a target trial, Figure \ref{fig:ehr-durable} depicts EHR-derived information for six non-surgical subjects from DURABLE, an NIH funded study examining long-term outcomes of bariatric surgery \citep{arterburn2020, coleman2016longterm, obrien2018microvascular, coleman2022bariatric, fisher2018association} using EHR data across three large health care organizations in the Kaiser Permanente system.

In each panel, information on hemoglobin A1c (A1c, blue dots) and body mass index (BMI, red dots) are plotted on the respective dates each measurement was recorded. BMI values are shown in relation to a cutoff of 35 kg/m$^2$, the typical clinical cutoff for consideration of bariatric surgery at Kaiser Permanente \citep{kaiserpermanente2023}, while A1c values are shown in relation to cutoffs of 5.7\% (prediabetes) and 6.5\% (diabetes) \citep{cdcA1c, diabetesADA}. Both BMI and A1c are important in ascertaining study eligibility when emulating hypothetical target trials \cite{obrien2018microvascular}. In simulations, for example, we consider two such studies of the effect of bariatric surgery on microvascular outcomes, the first of which has an eligibility criteria consisting of BMI ($\geq 35$ kg/m$^2$) and A1c ($\geq 5.7\%$) cutoffs, while the latter only requires a BMI exceeding $\geq 35$ kg/m$^2$. Our data application studies the same question in a population of individuals with diabetes and who have a BMI above $35$ kg/m$^2$, where both measures are necessary for determining the study eligible population.

Crucially, while some patients in Figure \ref{fig:ehr-durable} have rich data on both BMI and A1c (e.g. patients 1 and 4), others have rich information on BMI (e.g. 2, 3, and 5) or A1c only (e.g. patient 6). Furthermore, some have long gaps between measures of a certain type (e.g. patients 2, 3, 6) despite getting frequent measures of another type. Additionally, patients may frequently change which side of one of these clinically relevant cutoffs they belong to, and thus so too would their eligibility for inclusion in a particular study using one or more of these cutoffs as inclusion criteria.

\section*{Notation and Problem Set Up}\label{sec:setup}
\subsection*{Estimands of Interest}
Suppose interest lies in understanding the comparative effectiveness of some binary treatment on a time-to-event outcome, using data from an EHR database. Due to the rarity of subjects initiating exposure at any particular point in time, as is the case with bariatric surgery \citep{arterburn2014bariatric}, suppose that a sequence of trials are run and subsequently pooled together for analysis. 

To formalize this, let $m \in \{1, ..., M\}$ index a series of trials across $M$ regularly occurring intervals (e.g. months). Define $E_{mk} \in \{0,1\}$ to be a binary indicator denoting whether subject $k$ satisfies the eligibility criteria for inclusion in the study at baseline of trial $m$. Subjects may meet the study eligibility criteria at many points in time, and thus may be included in multiple trials. Let $A_{mk} \in \{0, 1\}$ denote the treatment status at baseline of trial $m$ for subject $k$, and $T_{mk} = \min(T^*_{mk}, T^C_{mk})$ be the observed event time for subject $k$ in trial $m$, which is the minimum of some true event time $T^*_{mk}$ and a censoring time $T^C_{mk}$.  Following Hernan et al. \cite{hernan2008, hernan2000marginal, hernan2016immortal}, we re-express outcomes on a discrete time scale as follows:

$$
\bm Y_{mk} = \begin{pmatrix} Y_{mk1} \\ Y_{mk2} \\ \vdots \\ Y_{mkT_{mk}}\end{pmatrix}~~\text{where}~~ Y_{mkt} = \begin{cases} 1 & T^*_{mk} \leq t \\ 0 & T^*_{mk} > t \end{cases}
$$

\noindent As will become clear, doing so enables the use of time-varying IPW to control for a variety of potential biases that cannot be easily addressed in continuous time \cite{hernan2000marginal}. 

Additionally, throughout, we let over bars denote variable history. For example, $\widebar Y_{mkt} = 0$ indicates that a subject has not experienced the outcome of interest through the end of the $t^{\text{th}}$ interval since baseline of trial $m$. Analogously, we denote treatment status for subject $k$ at time $t$ of trial $m$ as $A_{mkt}$, and treatment history $\widebar A_{mkt}$, with $\widebar A_{mkt} = a\mathds{1}_{t}$ denoting a constant treatment history of $a$ for all $t$ previous time periods.. Unlike when referencing outcome history, the subscript $t$ in exposure history (and later, covariate history) denotes the start of the $t^{\text{th}}$ interval. Finally, we let $Y^{(\widebar a)}_{mkt}$ denote counterfactual discrete survival indicators under treatment history $\widebar A_{mkt} = \widebar a$.

Possible estimands of interest are established using pooled logistic regression \citep{dagostino1990, hernan2000marginal, hernan2004structural}, typically with the goal of estimating a common effect across all time periods.

\begin{equation}\label{eq:target_estimand}
    \logit\biggr[P(Y^{(\widebar a)}_{mk(t+1)} = 1 | E_{mk} = 1, \widebar Y^{(\widebar a)}_{mkt} = 0, \widebar A_{mk(t+1)} = a\mathds{1}_{(t+1)})\biggr] = \psi^{(m)}_{0,t} + \psi a 
\end{equation}

In Equation \ref{eq:target_estimand}, $\psi$ represents the observational analogue of a per-protocol (PP) effect, with sustained adherence to baseline treatment strategy $A_{mkt} = A_{mk} = a$ for $0 \leq t \leq T_{mk}$. Interest could also lie in the observational analogue of an intention-to-treat (ITT) effect, that is the effect of recommendation to undergo treatment without requiring subsequent adherence. In practice, it is difficult to extract information from EHR data on treatment recommendations (i.e. doctor's suggestion to start or discontinue treatment) and as such, an ITT analogue might be proxied by replacing $\widebar A_{mk(t+1)} = a\mathds{1}_{(t+1)}$ in Equation \ref{eq:target_estimand} with $A_{mk} = a$ , thereby considering the effect of baseline treatment status only \citep{hernan2016target}. While such an estimand may be better referred to as a weak per-protocol effect, in contrast to the strong per-protocol effect of $A_{mkt} = A_{mk} = a$ for all $t$, we refer to it as an intention-to-treat effect as is typical in related literature \citep{hernan2008, hernan2016immortal, dickerman2019, yland2022, chiu2022, rossides2021, danaei2018, danaei2013observational}.

Notions of adherence in the context of our motivating bariatric surgery example are somewhat different than in other applications, where patients may frequently switch or stop treatment, perhaps due to adverse effects or clinician recommendations. Under the sequential TTE design, patients are routinely eligible for inclusion in multiple trials, and as such patients previously included in trials as a non-surgical control could undergo treatment and re-enter the study as an initiator of bariatric surgery at a later trial. For such subjects, all person-time following initiation of surgery is inconsistent with treatment ``assignment'' at baseline of these earlier trials in which they were considered controls, and in this sense, such person-time from the perspective of these earlier trials is non-adherent. While this technical distinction between per-protocol and ITT effects for bariatric surgery may be clinically less meaningful than in other applications, in part because we are treating bariatric surgery as non-reversible, and thus non-adherence as one sided, we keep this terminology for consistency with how such estimands are referenced in the literature on TTE \citep{hernan2008, hernan2016immortal, dickerman2019, yland2022, chiu2022, rossides2021, danaei2018, danaei2013observational}. Furthermore, we compute both estimands to illustrate use of this general framework which can be used in numerous EHR-based applications, including those where distinctions between per-protocol and ITT effects are clinically more meaningful. 

Under the marginal structural model in Equation \ref{eq:target_estimand}, $\psi$ denotes a causal log-odds ratio, which can be viewed as discrete hazard ratio (on the log scale) when prevalence of the outcome within any given time period is small \citep{dagostino1990, hernan2000marginal, robins2000marginal, dickerman2019}. $\psi^{(m)}_{0,t}$ is analogous to a baseline hazard at time $t$, which can vary by trial $m$. An alternative choice would be to estimate a shared baseline hazard across trials, $\psi_{0,t}$, but this may not be appropriate when the time period spanning all trials is large. Thus, while the baseline hazard is relative to time $t = 0$ (baseline) within each trial, this set-up allows for some form of heterogeneity across trials.

Critical to the definition of this model is the time-varying eligibility indicator $E_{mk}$, which is straightforward to regularly ascertain for some patients and impossible for other subjects (Figure \ref{fig:ehr-durable}). There are a number of ways in which missing eligibility criteria could arise and potentially result in bias. Figure \ref{fig:missing_DAGs} highlights three such ways which are explored in greater detail in subsequent simulation studies.

Finally, let $R_{mk}$ be a complete case indicator, that is whether individual $k$'s eligibility status can be ascertained at baseline of trial $m$. When subjects whose eligibility can not be ascertained are dropped from analysis, the estimands that standard methods will target implicitly change, specifically to:

\begin{equation}\label{eq:target_estimand_R}
    \logit\biggr[P(Y^{(\widebar a)}_{mk(t+1)} = 1 | E_{mk} = 1, R_{mk} = 1, \widebar Y^{(\widebar a)}_{mkt} = 0, \widebar A_{mkt} = a\mathds{1}_{(t+1)})\biggr] = \theta^{(m)}_{0,t} + \theta a 
\end{equation}

\noindent Selection bias occurs when $\theta \neq \psi$, that is, when the treatment effect among eligible complete cases does not equal the treatment effect among the entire eligible population. Note that the term selection bias is sometimes an umbrella term in the epidemiology literature and in related fields. For example, in Figure \ref{fig:missing_DAGs}, selection bias is used to describe bias which arises both from conditioning on a collider and dropping subjects from the population of interest when there is effect modification. While it's worth distinguishing that there are more than one way in which bias can arise from dropping subjects with incomplete eligibility criteria, our solution to all such sources of bias will the same, and as such in the remainder of the paper we simply refer to such biases as selection bias \citep{hernan2004structural}.

\subsection*{Notation for Additional Sources of Bias}
Selection bias is not the only potential source of bias in observational studies, and other common sources of bias, including confounding and differential censoring/non-adherence, require additional notation.  Despite the fact that treatment status $A_{mkt}$ may change over time, from the perspective of baseline of each trial $m$, we are only interested in treatment status as a point exposure, and thus time-varying confounding is not a concern (though baseline confounding remains a concern) \citep{robins2000marginal}.

Define $C_{mkt}$ to be a binary indicator denoting that subject $k$ was censored during time period $t$ of trial $m$ (that is $T^C_{mk} \in [t, t+1)$), and let $N_{mkt}$ represent a non-adherence indicator $\mathds{1}(A_{mk} \neq A_{mkt})$, denoting whether at time $t$ of trial $m$, subject $k$ has switched treatment from baseline (of trial $m$). $\widebar N_{mkt} = 0$ and $\widebar C_{mkt} = 0$ indicate that subject $k$ has remained adherent to baseline treatment status and uncensored, respectively, through the start of interval $t$ of trial $m$. 

Covariates will be necessary to address both selection bias and other sources of potential bias. We let $\bm L_{mk}$ be a vector of covariates at baseline of trial $m$, $\bm L_{mkt}$ a vector of covariates at the beginning of interval $t$ of trial $m$, and $\widebar{\bm L}_{mkt}$ covariate history. 

\section*{Methods}\label{sec:method}
\subsection*{Inverse Probability Weighting Framework}\label{sec:ipw_framework}

To address missing data in eligibility defining criteria, let baseline covariates $\bm L_{mk} = (\bm L_{mk}^e, \bm L_{mk}^c)$ where $\bm L^{e}_{mk}$ are the covariates which define eligibility for inclusion in trial $m$ and $\bm L_{mk}^c$ are all other baseline covariates. That is, $E_{mk} = g(\bm L^{e}_{mk})$ where $g(\cdot)$ is a deterministic eligibility defining rule. In previously conducted observational studies examining the long-term outcomes following bariatric surgery, eligibility rules have been functions of both A1c and/or BMI  \cite{arterburn2020, arterburn2015association, maciejewski2016, obrien2018microvascular, fisher2018association, coleman2016longterm}; we replicate this complex eligibility criteria in our simulations studies.

In the presence of missing eligibility data, we proceed by making the following novel eligibility missing at random (MAR) assumption:

\begin{equation}\label{eq:MAR}
R_{mk} \indep E_{mk}~|~\widebar{\bm L}^c_{mk}, \widebar{A}_{mk}, C_{mk} = 0    
\end{equation}

\noindent This MAR assumption states that whether or not a subject's eligibility status can be ascertained is independent of the value of their eligibility status, given histories for treatment and non-eligibility defining covariates. Utilizing Equation \ref{eq:MAR}, we propose weights for addressing potential selection bias due to the restriction of analysis to subjects with complete data.

\begin{equation}\label{ref:ipw_R}
W^{R}_{mk} = P(R_{mk} = 1~|~\widebar{\bm L}^c_{mk}, \widebar A_{mk},  C_{mk} = 0)^{-1}
\end{equation}

Even in the absence of missing eligibility data, care must be taken not to attribute differences in observed outcomes to differences in treatment status without accounting for the fact that treated subjects may be fundamentally different than unexposed subjects. In particular, confounding, differential loss to follow-up, and differential non-adherence must all be addressed in ways that allow rigorous comparison of these observed outcomes. IPW has previously been used in this analytical framework to address confounding and differential censoring, typically not distinguishing between loss to follow-up and non-adherence \citep{hernan2001marginal, hernan2008, dickerman2019, cain2010, danaei2013observational, petito2020, karim2018comparison}. Inverse probability weights which addresses these problems are as follows:

\begin{equation}
\begin{aligned}
W^{A}_{mk} &= P(A_{mk}~| ~\bm L_{mk},  E_{mk} = 1)^{-1} \\
W^C_{mkt} &= \prod_{i = 0}^t P(C_{mki} = 0 ~|~ \widebar C_{mk(i-1)} = 0, \widebar N_{mk(i-1)} = 0, \bm L_{mki}, E_{mk} = 1, A_{mk})^{-1}  \\
W^N_{mkt} &= \prod_{i = 0}^t P(N_{mki} = 0 ~|~ \widebar C_{mk(i-1)} = 0, \widebar N_{mk(i-1)} = 0, \bm L_{mki}, E_{mk} = 1, A_{mk})^{-1} 
\end{aligned}
\end{equation}

Working with observed data among subjects with ascertainable eligibility status amounts to fitting the following pooled logistic regression model:
\begin{equation}\label{eq:observed_PLR}
\logit\biggr[P(Y_{mk(t+1)} = 1 | E_{mk} = 1, R_{mk} = 1, \widebar Y_{mkt} = 0, A_{mk}, \widebar N_{mkt} = 0, \widebar C_{mkt} = 0)\biggr] = \tilde\theta^{(m)}_{0,t} + \tilde\theta A_{mk} 
\end{equation}

\noindent Fitting the model in Equation \ref{eq:observed_PLR} with weights 

\begin{equation}\label{ref:ipw}
W_{mkt} = W^{A}_{mk} \times W^{C}_{mkt} \times W^{N}_{mkt} \times W^{R}_{mk}
\end{equation}

\noindent recovers the parameter of interest $\psi$ when component weight models are correctly specified. Namely, weights $W^{A}_{mk} \times W^{C}_{mkt} \times W^{N}_{mkt}$ create a psuedo-population in which both treatment decisions and censoring are independent of covariates which may differ by treatment status (e.g. $\tilde\theta$ recovers $\theta$) \citep{robins2000marginal, hernan2004structural, hernan2024, dickerman2019, danaei2013observational, cain2010, karim2018comparison}. Similarly, weights $W^{R}_{mk}$ create a psuedo-population where eligibility status is independent of eligibility ascertainment (e.g. $\theta$ recovers $\psi$). Assumptions required for recovery of $\psi$ utilizing these inverse probability weights are discussed in the Supplementary Materials.

Without loss of generality, the model in Equation \ref{eq:observed_PLR} targets a per-protocol effect. Analysts choosing to pursue intention-to-treat effects could drop adherence ($\widebar N_{mkt} = 0$) from the model. In other settings, differential loss to follow-up may not be of concern. In either of these cases, Equation \ref{ref:ipw} can easily be modified to exclude irrelevant weights. Table \ref{tab:ipw_summary} summarizes component inverse probability weights, as well as their stabilized analogues, $SW$ \citep{hernan2024, karim2018comparison, hernan2001marginal, cain2010}.

In the remainder of the paper, we let $\hat\psi_{PP}$ and $\hat\psi_{ITT}$ denote estimates of observational analogues of per-protocol ($\psi_{PP}$) and ITT effects ($\psi_{ITT}$), respectively, from fitting the model in Equation \ref{eq:observed_PLR} with appropriate inverse probability weights.

\section*{Simulation Studies}
\label{sec:simulation_study}
In order to characterize realistic EHR-based settings under which missing eligibility data presents the possibility of selection bias and when it doesn't, and understand the finite sample performance of our proposed method, we developed a series of simulation studies. These simulations (sometimes known as plasmode simulations \cite{franklin2014plasmode}) are closely tied to data used to study bariatric surgery in DURABLE studies \citep{arterburn2020, coleman2022bariatric, coleman2016longterm, li2021fiveyear, fisher2018association}, and EHR data more generally. Briefly, the DURABLE database was used to both sample covariate vectors $\bm L_{mk}^c$ and estimate/inform simulation parameters for treatment, outcome, and time-varying eligibility models. Simulated treatments, outcomes, and eligibility statuses were then generated from these models and the covariates sampled. The complete EHR-based simulation infrastructure is described in detail in the Supplementary Materials.

Our simulations evaluated the performance of various analytical approaches under three missing data mechanisms for two different sets of eligibility criteria, one based on BMI ($\geq 35$ kg/m$^2$) and A1c ($\geq 5.7\%$) cutoffs (``hypothetical study \#1''), and one based only on the BMI cutoff (``hypothetical study \#2''). Due to the identical nature of eligibility for bariatric surgery and inclusion criteria for the second hypothetical study, eligibility ascertainment in this study was only an issue in the non-surgical arm. In both hypothetical studies, interest lay in the per-protocol effect of bariatric surgery on vascular disease outcomes \citep{fisher2018association, obrien2018microvascular}.

\subsection*{Missing Data Mechanisms}\label{sec:simulation_missingness}

\begin{figure}[h]
    \centering
      \begin{subfigure}[t]{0.3\linewidth}
        \begin{tikzpicture}[auto, node distance=1cm]
            \tikzstyle{square}=[rectangle, draw, minimum size=7mm]
            \tikzstyle{circ}=[circle, draw, minimum size=5mm]
            
            \node (LRA) {$\bm L^{RA}$};
            \node[above = of LRA] (LRY) {$\bm L^{RY}$};
            \node[above = of LRY] (LAY) {$\bm L^{AY}$};
            \node[right = of LRA] (A) {$A$};
            \node[right =of A] (Y) {$Y$};
            \node[square, below=  of LRA] (R) {$R$};
            \node[square, dashed, left = of R] (E) {$E$};
            \node[left = 0.7cm of LRY] (Le) {$\bm L^{e}$};

            \node[draw = red, fill = red, opacity = 0.05, inner sep=2mm, line width=.8pt,rounded corners, label=above:$\color{red}{\bm L^c}$,fit= (LRA) (LRY) (LAY)] {};

            \draw[ultra thick, ->] (Le) -- (E);
            \draw[ultra thick, ->] (A) -- (Y);
            \draw[violet, ultra thick, ->] (LRA) -- (R);
            \draw[violet, ultra thick, ->] (LRA) -- (A);
            \draw[violet, ultra thick, ->, bend left] (LRY) to (Y);
            \draw[violet, ultra thick, ->, bend right] (LRY) to (R);

            \draw[->, dashed, bend left] (LAY) to (A);
            \draw[->, dashed, bend left] (LAY) to (Y);
            \draw[->, dashed] (Le) to (A);
            \draw[->, dashed, bend left = 45] (Le) to (Y);
            \draw[->, dashed, ] (LRY) to (A);
            \draw[->, dashed, bend left] (LRA) to (Y);
    
        \end{tikzpicture}
    \caption{M-Bias}
    \label{fig:missing_DAGs_A}
    \end{subfigure} 
    \hfill
    \begin{subfigure}[t]{0.3\linewidth}
        \begin{tikzpicture}[auto, node distance=1cm]
            \tikzstyle{square}=[rectangle, draw, minimum size=7mm]
            \tikzstyle{circ}=[circle, draw, minimum size=5mm]
            
            \node (LRA) {$\bm L^{RA}$};
            \node[above = of LRA] (LRY) {$\bm L^{RY}$};
            \node[above = of LRY] (LAY) {$\bm L^{AY}$};
            \node[right = of LRA] (A) {$A$};
            \node[right =of A] (Y) {$Y$};
            \node[] at (2.7,-0.025) (M) {};
            \node[square, below=  of LRA] (R) {$R$};
            \node[square, dashed, left = of R] (E) {$E$};
            \node[left = 0.7cm of LRY] (Le) {$\bm L^{e}$};

            \node[draw = red, fill = red, opacity = 0.05, inner sep=2mm, line width=.8pt,rounded corners, label=above:$\color{red}{\bm L^c}$,fit= (LRA) (LRY) (LAY)] {};

            \draw[ultra thick, ->] (Le) -- (E);
            \draw[ultra thick, ->] (A) -- (Y);
            \draw[violet, ultra thick, ->] (LRA) -- (R);
            \draw[violet, ultra thick, ->] (LRA) -- (A);
            \draw[violet, ultra thick, ->, bend left] (LRY) to (Y);
            \draw[violet, ultra thick, ->, bend right] (LRY) to (R);
            \draw[violet, ultra thick, ->] (LRY) to (M.center);

            \draw[->, dashed, bend left] (LAY) to (A);
            \draw[->, dashed, bend left] (LAY) to (Y);
            \draw[->, dashed] (Le) to (A);
            \draw[->, dashed, bend left = 45] (Le) to (Y);
            \draw[->, dashed, ] (LRY) to (A);
            \draw[->, dashed, bend left] (LRA) to (Y);
    
        \end{tikzpicture}
    
    \caption{Treatment Effect~~~~~~\\Heterogeneity}
    \label{fig:missing_DAGs_B}
    \end{subfigure}
    \hfill
    \begin{subfigure}[t]{0.3\linewidth}
        \begin{tikzpicture}[auto, node distance=1cm]
            \tikzstyle{square}=[rectangle, draw, minimum size=7mm]
            \tikzstyle{circ}=[circle, draw, minimum size=5mm]
            
            \node (LRA) {$\bm L^{RA}$};
            \node[above = of LRA] (LRY) {$\bm L^{RY}$};
            \node[above = of LRY] (LAY) {$\bm L^{AY}$};
            \node[right = of LRA] (A) {$A$};
            \node[right =of A] (Y) {$Y$};
            \node[square, below=  of LRA] (R) {$R$};
            \node[square, dashed, left = of R] (E) {$E$};
            \node[left = 0.7cm of LRY] (Le) {$\bm L^{e}$};
            \node[circ, dashed, right = of R] (M) {$M$};

            \node[draw = red, fill = red, opacity = 0.05, inner sep=2mm, line width=.8pt,rounded corners, label=above:$\color{red}{\bm L^c}$,fit= (LRA) (LRY) (LAY)] {};

            \draw[ultra thick, ->] (Le) -- (E);
            \draw[ultra thick, ->] (A) -- (M);
            \draw[ultra thick, ->] (M) -- (Y);
            \draw[violet, ultra thick, ->] (LRA) -- (R);
            \draw[violet, ultra thick, ->] (LRA) -- (A);
            \draw[violet, ultra thick, ->, bend right] (LRY) to (R);
            \draw[violet, ultra thick, ->] (LRY) to (M);

            \draw[->, dashed, bend left] (LAY) to (A);
            \draw[->, dashed, bend left] (LAY) to (Y);
            \draw[->, dashed] (Le) to (A);
            \draw[->, dashed, bend left = 45] (Le) to (Y);
            \draw[->, dashed, ] (LRY) to (A);
            \draw[->, dashed, bend left] (LRA) to (Y);
    
        \end{tikzpicture}
    \caption{M-Bias Through Unobserved Mediator}
    \label{fig:missing_DAGs_C}
    \end{subfigure}
    \caption{Directed acyclic graphs (DAG) outlining missing data mechanisms used in simulations. Bold lines connecting nodes indicate relationships specified in the simulation study. Dashed lines indicate relationship pairs which could introduce confounding. While simulations did not introduce confounding, the dashed lines show that confounding can exist and be adjusted for without violation of missing at random assumption (Equation \ref{eq:MAR}). Finally, purple lines denote paths along which selection bias is introduced. For simplicity, each DAG represents a single time point $t$.}
    \label{fig:missing_DAGs}
\end{figure}
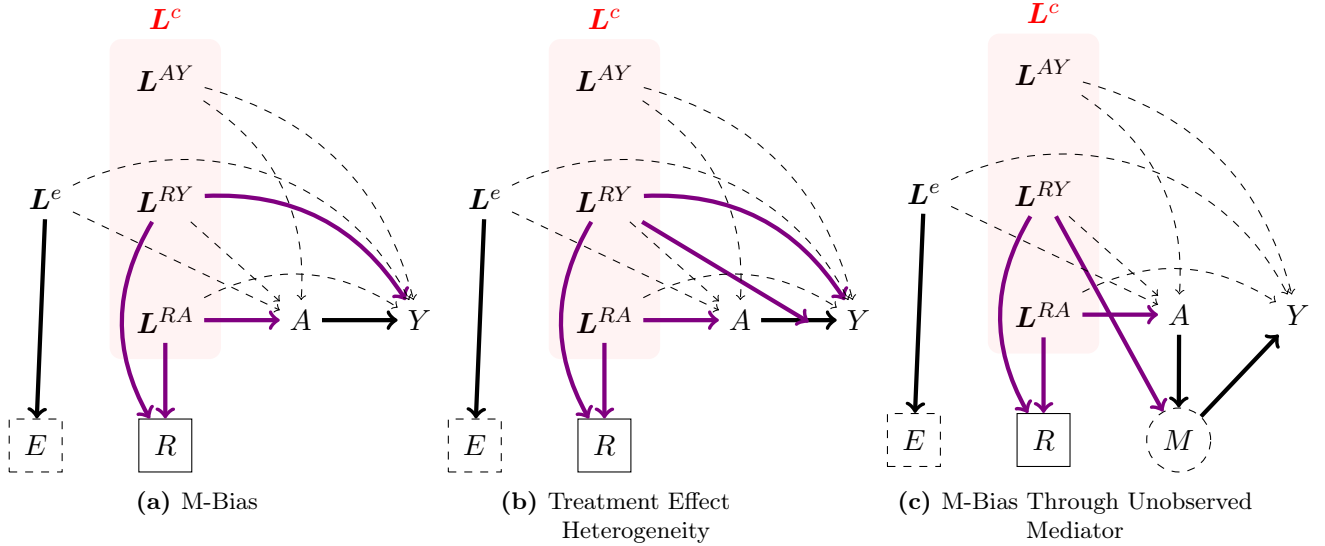

We utilized three mechanisms for generating missing eligibility criteria, outlined in Figure \ref{fig:missing_DAGs}. These mechanisms are not meant to be exhaustive, but rather are meant to demonstrate a few plausible structures by which selection bias may exist due to missing eligibility criteria. Under each missingness mechanism, complete case indicators $R_{mk}$ were generated from a logistic regression model using covariates within $\bm L_{mk}^c$, some of which might also be used in models for sampling treatment assignments or outcomes, depending on the setting. For simulated subjects with $R_{mk} = 0$, eligibility defining covariates $\bm L_{mk}^e$, and thus so too eligibility status $E_{mk}$, were set to be missing.

To elaborate on the specific missing data mechanisms explored, let $\bm L^c = (\bm L^{AY}, \bm L^{RA}, \bm L^{RY})$ where $\bm L^{AY}$ denotes traditional confounders (i.e. covariates associated with both treatment and outcome), $\bm L^{RA}$ denotes covariates associated with both eligibility ascertainment ($R$), and treatment, and $\bm L^{RY}$ denotes covariates associated with both eligibility ascertainment and outcome. Figure \ref{fig:missing_DAGs_A} demonstrates M-bias, also referred to as collider bias, whereby one conditions on a common effect of a cause of treatment $(\bm L^{RA})$ and a cause of outcome $(\bm L^{RY})$. Figure \ref{fig:missing_DAGs_B}  extends this structure to include treatment effect heterogeneity in the form of an interaction between $A$ and some component(s) of $\bm L^{RY}$. Finally, Figure \ref{fig:missing_DAGs_C}  considers a more complex type of M-bias where $\bm L^{RY}$ is not directly associated with outcome but instead is associated with a mediator $M$ through which treatment $A$ is associated with outcome $Y$. In the context of these simulations, where interest lies in vascular outcomes, examples of such mediators are BMI and A1c values, whose respective frequency of observation is highly heterogeneous.

In practice, these sets of variables are unlikely to be disjoint, especially in EHR-based settings where certain covariates are likely associated with eligibility ascertainment, treatment and outcome simultaneously. Nevertheless, we find it useful to make this distinction explicit to demonstrate distinct ways in which selection bias may manifest itself in this setting in the absence of confounding, a challenge for which extensive literature exists \citep{hernan2024, hernan2001marginal, hernan2000marginal, schneeweiss2005review, sturmer-adjusting-2004, sturmer-performance-2007, toh-confounding-2011, brookhart2010instrumental}. Paths by which confounding could exist are represented by dotted arrows in Figure \ref{fig:missing_DAGs}.

\subsection*{Analytical Methods}
\label{sec:simulation_analytical_methods}

In order to understand finite sample properties of various estimators of $\psi_{PP}$, we computed the model in Equation \ref{eq:observed_PLR} with different combinations of weights $W^{R}_{mk}$ and $W^N_{mkt}$. Confounding was not considered in simulations, as the use of IPW to address confounding is well understood. For analogous reasons, differential right-censoring (e.g. due to loss to follow-up) was not considered. Thus, weights for confounding ($W^{A}_{mk}$) and differential censoring ($W^{C}_{mkt}$) were not required for analysis of simulated datasets.

All component weights were estimated using pooled logistic regression. Specifications for selection bias weights included modeling $R$ with $\bm L^{RA}$ and $\bm L^{RY}$ each individually (incomplete adjustment via omission of certain variables), $\bm L^{RA}$ and $\bm L^{RY}$ together (denoted $\bm L^R$; correct model specification), and $\bm L^R$ and $A$ (inclusion of $A$ superfluous). Non-adherence weights, when utilized, were correctly specified and included all covariates which influenced treatment decisions (denoted $\bm L^A$). Equation \ref{eq:observed_PLR} was estimated using each possible specification of weights $W_{mk} = W^{R}_{mk} \times W^N_{mkt}$, as well as combinations where either or both component weights were omitted entirely. 

In hypothetical study \# 2, where eligibility criteria comprised only of a BMI measurement of 35 kg/m$^2$ or greater, we also evaluated estimating selection bias weights from separate models stratified by treatment status. Due to the fact that eligibility for inclusion in this hypothetical study (at the time of treatment initiation) was guaranteed by electing to undergo bariatric surgery, $P(R_{mk} ~|~ A_{mk} = 1) = 1$ and thus so too should weights $W^{R}_{mk}$ in order to guarantee correct specification.

\subsection*{Simulation Results}
\label{sec:simulation_results}

Results from hypothetical study \# 1 are shown in Table \ref{table:results_1} and results from hypothetical study \# 2 are shown in Table \ref{table:results_2}. In all scenarios, estimation of $\psi_{PP}$ without inverse probability of selection weights resulted in bias ranging from 7-188\%. Correct specification of $W^R_{mk}$ recovered unbiased estimation. In fact, in most settings, incorrect specification of $W^R_{mk}$ via omission of $\bm L^{RA}$ still performed reasonably well, suggesting that the association between selection and outcome may drive selection bias more than the association between selection and treatment. 

In the second study, correct weight specification required estimating separate selection bias weights for those who were treated and untreated. Fitting a single model to all subject-trials regardless of treatment status, even one with the correct functional form within the untreated arm, introduced significant bias in $\hat\psi_{PP}$. This is likely because in hypothetical study \# 2, true weights $W_{mk}^R = 1$ for treated individuals and incorrectly including such subjects when fitting a single selection bias weight model biased estimation of weights for all individuals. Altogether, this suggests that if patterns of eligibility ascertainment are believed to differ substantially between treatment arms, a single parametric model may be insufficient for capturing how $R_{mk}$ depends on $\bar L^c_{mk}$ and $\bar A_{mk}$ when modeling selection bias weights $W^R_{mk}$.

\section*{Data Application}
\label{sec:application}
To illustrate an application of our IPW framework, with a particular focus on how researchers might think about missing eligibility data in a real EHR example, we studied the long term effect of bariatric surgery on microvascular outcomes, a composite outcome consisting of nephropathy, neuropathy, and retinopathy, among patients with Type 2 diabetes mellitus (T2DM) using a sequential TTE design. Using the DURABLE database, O'Brien and colleagues previously examined this question under a matched cohort design \cite{obrien2018microvascular}. In this previous work, the authors used data up to two years prior to index date (e.g. the date of surgery for the surgical case in each matched group) to ascertain eligibility, excluding patients whose eligibility status could not be determined due to missing data. Determining how far to look back when attempting to ascertain eligibility can be viewed as a bias-variance trade off, whereby longer lookback windows will enable ascertainment of more subjects' eligibility status, but some subjects will be deemed eligible or ineligible incorrectly. The decision to look back two years isn't necessarily right or wrong, but is a distinct choice made by the authors, whose influence on results we explored in this re-analysis. In addition to differences in study design and eligibility lookbook windows, we report a common hazard ratio (eg. averaged over follow-up time) in contrast to time-varying estimates reported by O'Brien et al. This decision was made in part for computational efficiency, given that analysis was re-run over several combinations of eligibility lookback windows. Other design aspects including study eligibility criteria, follow-up time, treatment and outcome definitions, and reasons for censoring subjects follow immediately from O'Brien et al. \citep{obrien2018microvascular}

\subsection*{Overview of Data}
\label{sec:data_overview}
Analysis was restricted to patients with diabetes eligible for bariatric surgery between January 1, 2005 and December 31, 2011, with follow-up allowed until September 30, 2015. Under the sequential TTE design, this yielded $M = 84$ trials, beginning at the start of each month between January 1, 2005 and December 31, 2011. Eligibility criteria consisted of a BMI $\geq 35$ kg/m$^2$ and diagnosis of T2DM. Using a similar definition to that of O'Brien et al., T2DM was defined as having the most recent A1c measurement $\geq$ 6.5\% or fasting blood glucose measurement $\geq 126$ mg/dL, within a pre-specified lookback period before the start of trial $m$, or by having an active prescription for a diabetes medication when trial $m$ began. Exclusion criteria removed patients who had previously undergone bariatric surgery, had a history of microvascular disease, or had been pregnant in the previous 12 months. The endpoint of interest was time to incident microvascular disease (e.g. the first instance of nephropathy, neuropathy, or retinopathy). Patients were censored at the time of death, disenrollment from their health plan, cancer, following a period of 13 months without weight or blood pressure measurements, or end of study.

\subsection*{Ascertainment of Eligibility}
\label{sec:data_elig}

\begin{figure}[H]
    \centering
    \includegraphics[width = 0.92\textwidth]{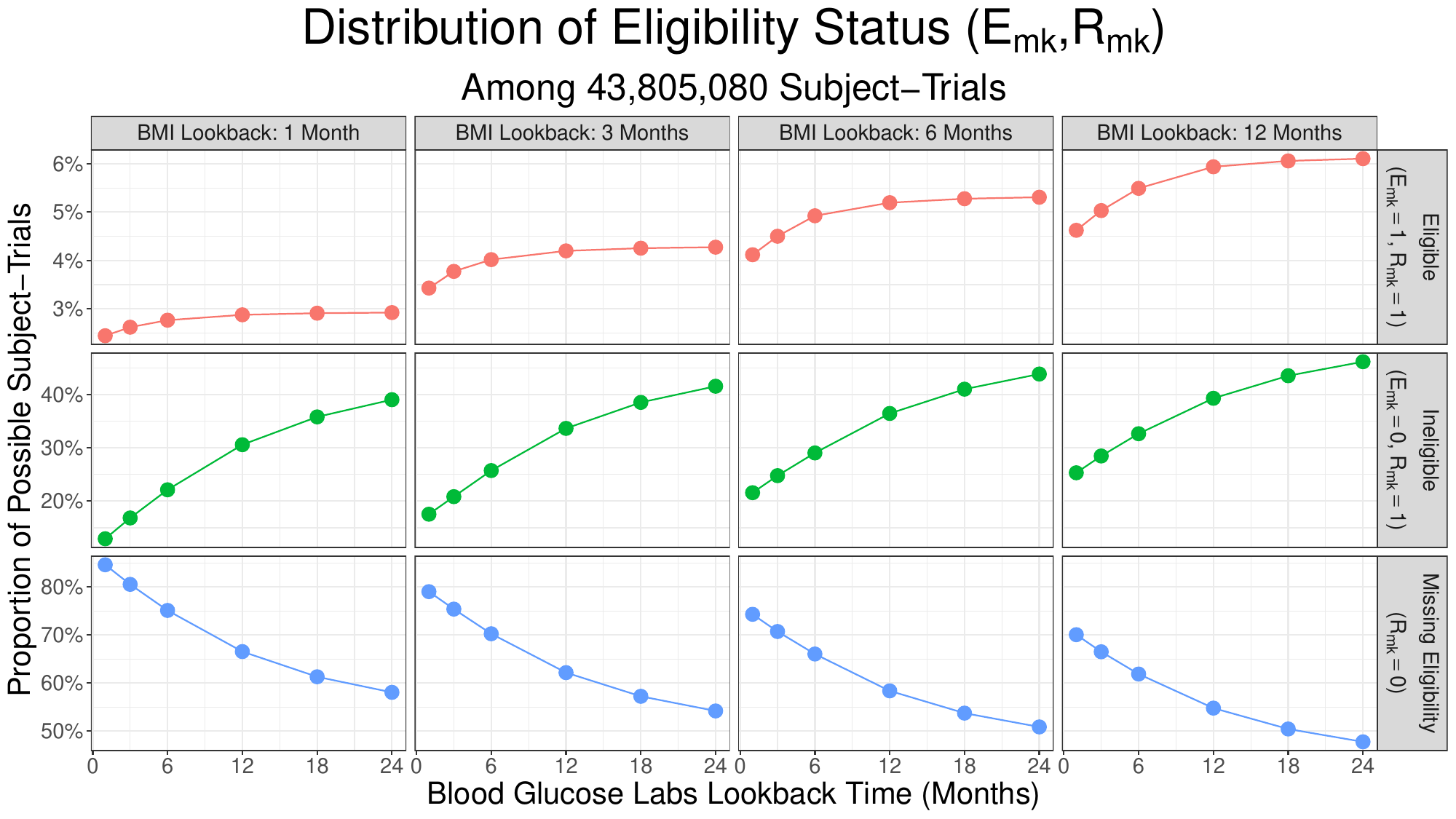}
    \caption{Joint distribution of eligibility and ascertainment status $(E_{mk}, R_{mk})$ across nearly 44 million subject-trials. In general, allowing longer lookback times before the start of trial $m$ increases the number of subjects for whom eligibility can be ascertained $(R_{mk} = 1)$. However, measurement values from longer lookbacks may less accurately reflect a subjects' eligibility defining covariates (as in Figure \ref{fig:ehr-durable}). Essentially, the choice of lookback windows can be framed as a sort of bias-variance trade off.}
    \label{fig:elig_dist}
\end{figure}

After filtering the data to exclude subject-trials where patients had a history of microvascular complications or were censored for one of the reasons mentioned above, 43,805,050 subject-trials across $K = 1,152,227$ subjects remained for which eligibility needed to be assessed. To study the sensitivity of results to the length of time used to establish study eligibility, we used a grid of lookback times. BMI was assessed using the most recent value within \{1, 3, 6, 12\} months before the start of trial $m$ and TD2M status was assessed using the most recent blood glucose lab measurements available with \{1, 3, 6, 12, 18, 24\} months. 

If measurements were unavailable for either of the two eligibility criteria, eligibility status was deemed missing ($R_{mk} = 0$). The distribution of eligibility status is shown in Figure \ref{fig:elig_dist}. Under the shortest lookback windows, eligibility could not be ascertained for over 80\% of subject-trials. Increasing lookback windows for both BMI and A1c increased the number of subjects for whom eligibility could be ascertained, though the number of subject trials with confirmed eligibility $(E_{mk} = 1, R_{mk} = 1)$ tended to plateau around the 12 month blood glucose lookback window.

\begin{figure}[H]
    \centering
    \includegraphics[width=0.8\textwidth]{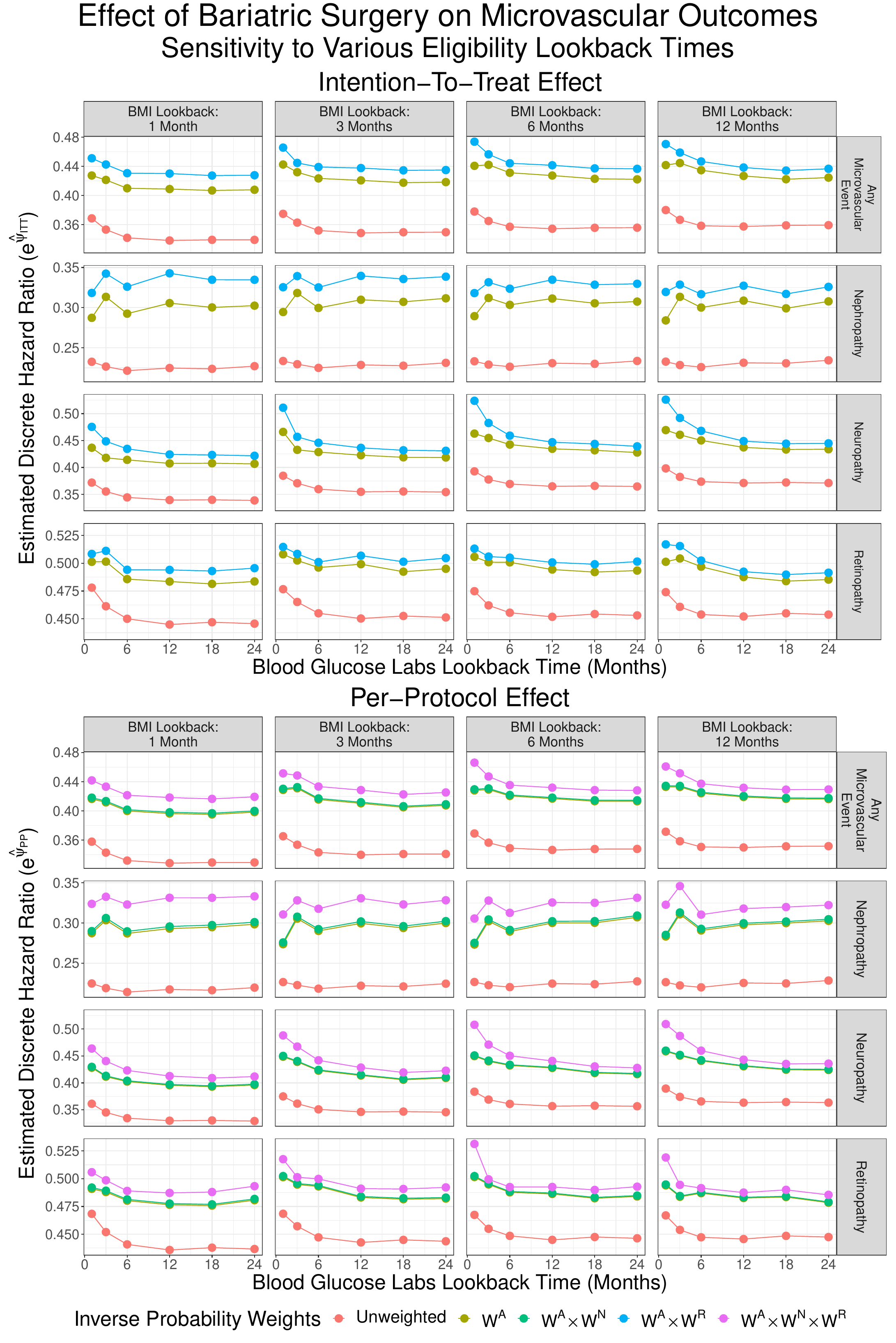}
    \caption{Intention-to-treat and per-protocol discrete hazard ratio estimates for the effect of bariatric surgery on microvascular outcomes. Eligibility ($E_{mk}$) was assessed over a grid of lookback windows  before the start of trials $m$ for both blood glucose and BMI measurements. Estimates are reported using various combinations of inverse probability weights. Note that Y-axis scales differ between component outcomes in order to highlight differences between various weight combinations, which frequently are smaller than those across component outcomes.}
    \label{fig:effect-sensitivity}
\end{figure}

\subsection*{Sensitivity of Effect Estimates to Eligibility Lookback Times}
\label{sec:data_sens}

ITT and per-protocol effect estimates for bariatric surgery on each individual outcome of interest, as well as the composite, are shown in Figure \ref{fig:effect-sensitivity}. Accounting for confounding through the inclusion of $W_{mk}^A$ attenuated effect estimates towards the null. Notably, accounting for the possibility of selection bias through inclusion of $W_{mk}^R$ also attenuated effect estimates, albeit to a lesser extent. 

Such findings are consistent with discussion of missing data by O'Brien and colleagues, who noted that subjects dropped for missing data were more likely to have had shorter duration of observed diabetes than subjects retained with complete data \cite{obrien2018microvascular}. One possible explanation for the smaller degree of attenuation following adjustment for selection bias than for confounding/non-adherence alone is that the covariates used to adjust for selection bias were a strict subset of the covariates used to address confounding/non-adherence. This need not be the case in all studies however.  

While it may not qualitatively change the fact that bariatric surgery is associated with long term decrease in risk of microvascular complications, failure to account for the possibility of selection bias due to missing eligibility data may lead one to overstate effect sizes by up to 10\%. In general, effect estimates were somewhat sensitive to choices of lookback times for both BMI and blood glucose lab measurements particularly for individual component outcomes. In some cases, differences between effect estimates using the longest and shortest lookback windows were on par with differences between estimates accounting and failing to account for selection bias.

\subsection*{Final Results}
\label{sec:data_results}
We picked a BMI lookback window of 3 months and a blood glucose lookback window of 12 months for which to report full 95\% confidence intervals using $B = 1,000$ bootstrap replicates resampled at the subject $(k)$ level. Confidence intervals were computed using a Normal approximation interval rather than a percentile-based interval method \cite{wasserman2004all, efron1993introduction} due to the limit of replicates that is computationally feasible on such a large dataset \cite{dickerman2022, hajage2022extracorporeal}. Inferential procedures for our IPW framework, including coverage properties of various bootstrap techniques, are discussed in the Supplementary Materials. 

Effect estimates from Table \ref{table:data_app} are not directly comparable to results reported by O'Brien and colleagues, who estimate time-varying hazard ratios HR$(t)$ as opposed to a common hazard ratio which we report \cite{obrien2018microvascular}. Nevertheless, results between our work and O'Brien et al. are qualitatively very similar, with estimates reported in this work perhaps slightly attenuated compared to what one might reasonably posit a common hazard ratio estimated by O'Brien et al. would be based on averaging HR$(t)$ over various reported time points. 

Despite utilizing a product of inverse probability weights, which are known to increase variance of associated estimators \citep{metten2022inverse, hernan2024}, 
efficiency of out reported estimates met or exceeded that reported by O'Brien et al. at any time point. This likely owes to the fact that our choice of design did not utilize matching, thereby retaining a much greater percentage non-surgical patients than the matched cohort design used by O'Brien and coauthors.

\section*{Discussion}
\label{sec:discussion}

The target trial emulation framework of Hernán and Robbins \cite{hernan2016target} has exploded in popularity in recent years and been lauded for laying out a road map which enables, or even forces researchers to think critically about complex design aspects through which bias may enter their study. Unfortunately, missing data, particularly missing data in eligibility defining criteria, is not part of this road map. In this work, we demonstrated that improper handling of missing eligibility data posed the threat of selection bias across a range of common scenarios in EHR-based observational studies, in hopes of demonstrating that consideration of missing eligibility data should be part of this road map going forward.

Crucial to both thinking about and accounting for missing eligibility criteria is the missing at random assumption postulated in Equation \ref{eq:MAR}. At its core, this MAR assumption states that whether or not one's eligibility status can be ascertained is independent of what that eligibility status is, after accounting for everything observable for all patients. As is typical of MAR assumptions, this MAR assumption is not testable in practice. Common violations of this assumption are likely to be found in cases where study eligibility criteria can be viewed as some notion of patient health not captured by covariates included in the analysis, with a patient's health driving how frequently they're followed in the EHR, a phenomenon similar to Berkson's bias \citep{westreich2012}. Examples of such eligibility criteria that implicitly reflect patient health and drive how frequently a subject is followed by their clinical care team might include CD4 cell count in the study of HIV progression, and glucose filtration rate (GFR) in the study of chronic kidney disease. In such hypothetical cases, sicker patients would likely be overrepresented in the analysis population in a manner not directly addressable by our IPW method, as the factor driving the missingness would itself be missing. Similar violations may also be observed when certain patients are more engaged in care that produces more frequent health care interactions and thus frequent measurements. Psychotherapy, for example, might specifically lead to gathering more information on mental health symptoms less likely to be collected in routine care that are desired eligibility criteria in some retrospective EHR-based observational study. 

When analysts feel this MAR assumption is violated, alternative MAR assumptions can be conceived. For example, one could additionally condition on things like eligibility history (both status $E$ and eligibility defining covariates $\bm L^e$). This relaxed assumption may be more likely to hold at the expense of presenting analysts with non-monotone missingness, something much harder to rectify. While violations of this MAR assumption are not ideal, violations where subjects with missing eligibility data were less likely to have been eligible seem somewhat less harmful than the other direction, where subjects without eligibility defining covariates were more likely to have been eligible.

We noted that one possible explanation why adjusting for selection bias only attenuated effect estimates toward the null by a small amount in our data application was that the set of variables used to adjust for these two sources of bias were very similar. While the focus of this work was on marginal effect estimates, such an observation leads us to hypothesize that the possibility of selection bias may not be as bad if conditional effect estimates are of primary interest and the set of covariates associated with selection, treatment, and outcome are similar. Nevertheless, one still needs to consider the possibility of selection bias in this setting, as Tompsett et al. showed that selection bias was still a concern when one estimated marginal effects by averaging over conditional effect estimates under a linear regression model \cite{tompsett2023target}.

Tompsett et al. discussed a notion of MAR related to eligibility criteria but never provided a formal statement of what a MAR assumption would look like in their framework. In contrast to an IPW approach, Tompsett et al. proposed multiple imputation of eligibility defining criteria, something they demonstrated can also address selection bias. This approach may not be feasible in some EHR-based observational studies, particularly when a sequence of target trials are pooled together. For example, if we had attempted to apply multiple imputation in our data application, we'd be attempting to impute eligibility status for 50-85\% of subject trials (Figure \ref{fig:elig_dist}). In their supplementary materials, Tompsett and colleagues note inverse probability weighted outcome regressions as an alternative solution, but found limited improvement for this estimator compared to unweighted outcome regressions. Combinations of IPW and imputation have been used to address missing data in other contexts \citep{thaweethai2021robust, seaman2012combining}, and may offer a flexible path forward for dealing with non-monotone missingness in eligibility defining covariates as well as simultaneous missingness in other variables such as confounders or outcomes. 

We pursued an IPW approach for several reasons. To begin with, it's a natural choice given that other concerns are already commonly addressed using IPW, so inclusion of additional weights fits seamlessly into existing analysis pipelines. Furthermore, addressing missing data with IPW allows resulting models to recover whatever effect estimates they would target under complete data \citep{breslow1999}. Finally, notions of multiply robust estimators \cite{bang2005doubly} for addressing confounding, censoring, and non-adherence are not typically utilized in TTE studies even when missing eligibility data is not a concern \cite{scola2023implementation, hoffman2022}, so addressing these challenges plus missing data in a more robust manner than IPW seems both substantially more difficult and incongruent to methods used in practice.

One of the first steps in any retrospective EHR-based observational study is to clearly articulate the eligible study population to which the results are intended to speak to, or generalize. Thus for any retrospective EHR-based  observational study in which missing data may preclude characterizing complete eligibility criteria, it is plausible that selection bias could arise. It is impossible to give a general rule of thumb for the degree to which selection bias may or may not be an issue for any arbitrary study design or causal questions. This will depend on the relative availability of data, as well as factors which drive treatment, outcome, and missingness -- factors which may vary greatly by study and question.

Given the need to ascertain eligibility for subjects repeatedly throughout time, the problem of selection bias is possibly of greater concern in the sequential TTE design than in other study designs. Nevertheless, selection bias due to missing eligibility criteria remains a possible worry in other designs, and the IPW solutions explored in the work might apply in such cases. In non-sequential TTE and other designs where time-zero is naturally defined by treatment initiation, eligibility status need only be queried at a single time point. As such, weights would simplify to the form  $W^R_{k}$, as only a single weight is necessary per subject at baseline. Further work may be necessary to extend the ideas of this framework to designs where patients can be meet study eligibility criteria at multiple points in time but may only enter the study at a single time point. One such example is the matched cohort design, which has been used as alternative to sequential TTE to establish start of follow-up for initiation of no treatment, particularly in the study of long term outcomes following bariatric surgery \citep{obrien2018microvascular, courcoulas2013weight, arterburn2020, coleman2016longterm, coleman2022bariatric, fisher2018association, koffman_evaluation2021, Harrington2024}.

Though this work focused on comparisons between initiators and non-initiators of treatment, similar problems and solutions may exist in other applications comparing head-to-head treatments (active comparator new user design). Selection bias due to missing eligibility criteria may be less of a concern in certain active comparator new user designs in which study and treatment eligibility are identical and initiation of treatment implies study eligibility, as was the case for bariatric surgery in hypothetical study \# 2 in our simulations.  If selection bias exists in such cases, it is more likely to be non-differential between the two treatment arms, potentially causing less bias when comparing outcomes between the two active comparator groups. Additional work may be necessary for studies interested in comparisons of prevalent and new users of a particular treatment, which seems related to certain matched cohort designs in that patients can be meet study eligibility criteria at multiple points in time but may only enter the study at a single time point \citep{suissa2017}.

In order to tackle the challenge of selection bias due to missing eligibility criteria, the assumption that missing data is restricted to eligibility defining covariates is implicit throughout this work. Such an assumption is unlikely to hold in practice when using EHR databases, and missing data can affect ascertainment of confounders, treatment, and outcomes, perhaps simultaneously. Related to missing data is the possibility of misclassification, arising from the assumption that the absence of a procedure or diagnostic code implies the absence of a condition. This very assumption was used to establish initiation of bariatric surgery as well as component microvascular outcomes in our data application. Though a possible limitation of this work, greater literature exists for these challenges \cite{levis2022, williamson2012, sun2018, perkins2018, harel2018, Funk2014}, which might be combined with the solutions outlined in this work for missing eligibility criteria. For example, the use of imputation for components of $\bm L^c$ within our data application is discussed in greater detail in the Supplementary Materials.

Finally, a reasonable critique of this work is the need to get multiple weight models correctly specified to obtain consistent estimation of effects of interest. Analysts may be able to get reasonably close to correct models but are likely to misspecify some or all weight models in practice. Thus, a goal of future work is to look towards semi-parametric theory \citep{levis2022, kennedy2023semiparametric, tsiatis_semiparametric_2006} with the goal of establishing robust and efficient estimators which perform well even under some degree of model misspecification.

\section*{Code Availability}
All code for analysis and simulations is made available on GitHub at \url{https://github.com/lbenz730/missing_eligibility_tte}.

\section*{Funding}
This work was supported by NIH Grants R01 DK128150-01 (LB, RM, RW, DA, CL, HF, SMS, SH) and F31 DK141237-01 (LB).

\clearpage

\bibliographystyle{ama}
\bibliography{ref}

\newpage

\begin{table}[htb]
    \centering
    \begin{tabular}{|Sc|Sc|Sc|}
        \hline
        \textbf{Weight}  & \textbf{Purpose} & \textbf{Definition} \\
        \hline
         $W^{A}_{mk}$ & Confounding  & $P(A_{mk}~| ~\bm L_{mk},  E_{mk} = 1)^{-1}$ \\
         $W^{C}_{mkt}$ & Censoring  & $\prod_{i = 0}^t P(C_{mki} = 0 ~|~ \widebar C_{mk(i-1)} = 0, \widebar N_{mk(i-1)} = 0, \bm L_{mki}, E_{mk} = 1, A_{mk})^{-1} $ \\
         $W^{N}_{mkt}$ & Non-Adherence  & $\prod_{i = 0}^t P(N_{mki} = 0 ~|~ \widebar C_{mk(i-1)} = 0, \widebar N_{mk(i-1)} = 0, \bm L_{mki}, E_{mk} = 1, A_{mk})^{-1} $ \\
         $W^{R}_{mk}$ & Selection Bias  & $P(R_{mk} = 1~|~\widebar A_{mk}, \widebar{\bm L}^c_{mk}, C_{mk} = 0)^{-1}$ \\
         \hline
         $SW^{A}_{mk}$ & Confounding  & $\frac{P(A_{mk}~|~E_{mk} = 1)}{P(A_{mk}~| ~\bm L_{mk},  E_{mk} = 1)}$ \\
         $SW^{C}_{mkt}$ & Censoring  & $\frac{\prod_{i = 0}^t P(C_{mki} = 0 ~|~ \widebar C_{mk(i-1)} = 0, \widebar N_{mk(i-1)} = 0, E_{mk} = 1, A_{mk})}{\prod_{i = 0}^t P(C_{mki} = 0 ~|~ \widebar C_{mk(i-1)} = 0, \widebar N_{mk(i-1)} = 0, \bm L_{mki}, E_{mk} = 1, A_{mk})}$ \\
         $SW^{N}_{mkt}$ & Non-Adherence  & $\frac{\prod_{i = 0}^t P(N_{mki} = 0 ~|~ \widebar C_{mk(i-1)} = 0, \widebar N_{mk(i-1)} = 0, E_{mk} = 1, A_{mk})}{\prod_{i = 0}^t P(N_{mki} = 0 ~|~ \widebar C_{mk(i-1)} = 0, \widebar N_{mk(i-1)} = 0, \bm L_{mki}, E_{mk} = 1, A_{mk})}$ \\
         $SW^{R}_{mk}$ & Selection Bias  & $\frac{P(R_{mk} = 1~|~\widebar A_{mk}, C_{mk} = 0)}{P(R_{mk} = 1~|~\widebar{\bm L}^c_{mk}, \widebar A_{mk}, C_{mk} = 0)}$ \\
         \hline
    \end{tabular}
    \caption{Summary of component inverse probability weights}
    \label{tab:ipw_summary}
\end{table}

\begin{table}[H]
\scriptsize
\centering
\begin{tabular}{|>{}Sc|Sc|Sc|Sc|Sc|Sc|c|>{}c|}
\hline
\multicolumn{2}{|c|}{\textbf{Setting}} & \multicolumn{2}{c|}{\textbf{IPW Models}} & \multicolumn{2}{c|}{\textbf{Mean} $\bm{ \hat \psi}_{PP}$} & \multicolumn{2}{Sc|}{\textbf{Median} $\bm{ \hat{ \psi}_{PP}}$} \\
\cline{1-2} \cline{3-4} \cline{5-6} \cline{7-8}
\textbf{Missingness} & $\bm\psi_{PP}$ & $\bm W_{mk}^{R}$ & $\bm W_{mkt}^{N}$ & \textbf{Bias} & \textbf{\% Bias} & \textbf{Bias} & \textbf{\% Bias}\\
\hline
 &  &  & --- & -0.021 & 6.7 & -0.023 & 7.2\\
\cline{4-8}
 &  & \multirow{-2}{*}{\centering\arraybackslash ---} & $N \sim \bm{L}^{A}$ & -0.020 & 6.3 & -0.022 & 6.7\\
\cline{3-8}
 &  &  & --- & -0.012 & 3.8 & -0.009 & 2.7\\
\cline{4-8}
 &  & \multirow{-2}{*}{\centering\arraybackslash $R \sim \bm{L}^{RA}$} & $N \sim \bm{L}^{A}$ & -0.020 & 6.2 & -0.018 & 5.7\\
\cline{3-8}
 &  &  & --- & -0.010 & 3.2 & -0.016 & 5.1\\
\cline{4-8}
 &  & \multirow{-2}{*}{\centering\arraybackslash $R \sim \bm{L}^{RY}$} & $N \sim \bm{L}^{A}$ & -0.009 & 2.9 & -0.016 & 5.0\\
\cline{3-8}
 &  &  & --- & 0.011 & -3.4 & 0.011 & -3.3\\
\cline{4-8}
 &  & \multirow{-2}{*}{\centering\arraybackslash $R \sim \bm{L}^{R}$} & $N \sim \bm{L}^{A}$ & 0.002 & -0.7 & 0.002 & -0.5\\
\cline{3-8}
 &  &  & --- & 0.011 & -3.4 & 0.011 & -3.3\\
\cline{4-8}
\multirow{-10}{*}{\centering\arraybackslash M-Bias} & \multirow{-10}{*}{\centering\arraybackslash -0.322} & \multirow{-2}{*}{\centering\arraybackslash $R \sim \bm{L}^{R} + A$} & $N \sim \bm{L}^{A}$ & 0.002 & -0.7 & 0.000 & 0.0\\
\cline{1-8}
 &  &  & --- & 0.036 & -11.7 & 0.040 & -13.3\\
\cline{4-8}
 &  & \multirow{-2}{*}{\centering\arraybackslash ---} & $N \sim \bm{L}^{A}$ & 0.036 & -11.7 & 0.039 & -12.9\\
\cline{3-8}
 &  &  & --- & 0.035 & -11.4 & 0.034 & -11.1\\
\cline{4-8}
 &  & \multirow{-2}{*}{\centering\arraybackslash $R \sim \bm{L}^{RA}$} & $N \sim \bm{L}^{A}$ & 0.034 & -11.2 & 0.035 & -11.6\\
\cline{3-8}
 &  &  & --- & -0.005 & 1.7 & -0.002 & 0.6\\
\cline{4-8}
 &  & \multirow{-2}{*}{\centering\arraybackslash $R \sim \bm{L}^{RY}$} & $N \sim \bm{L}^{A}$ & -0.005 & 1.7 & -0.001 & 0.4\\
\cline{3-8}
 &  &  & --- & -0.004 & 1.3 & 0.003 & -1.0\\
\cline{4-8}
 &  & \multirow{-2}{*}{\centering\arraybackslash $R \sim \bm{L}^{R}$} & $N \sim \bm{L}^{A}$ & -0.005 & 1.6 & 0.003 & -1.1\\
\cline{3-8}
 &  &  & --- & -0.004 & 1.3 & 0.003 & -1.1\\
\cline{4-8}
\multirow{-10}{*}{\centering\arraybackslash \shortstack{Treatment\\Effect\\Heterogeneity}} & \multirow{-10}{*}{\centering\arraybackslash -0.305} & \multirow{-2}{*}{\centering\arraybackslash $R \sim \bm{L}^{R} + A$} & $N \sim \bm{L}^{A}$ & -0.005 & 1.6 & 0.003 & -1.1\\
\cline{1-8}
 &  &  & --- & 0.057 & -7.0 & 0.063 & -7.7\\
\cline{4-8}
 &  & \multirow{-2}{*}{\centering\arraybackslash ---} & $N \sim \bm{L}^{A}$ & 0.057 & -6.9 & 0.062 & -7.5\\
\cline{3-8}
 &  &  & --- & 0.056 & -6.8 & 0.061 & -7.4\\
\cline{4-8}
 &  & \multirow{-2}{*}{\centering\arraybackslash $R \sim \bm{L}^{RA}$} & $N \sim \bm{L}^{A}$ & 0.055 & -6.7 & 0.061 & -7.5\\
\cline{3-8}
 &  &  & --- & -0.001 & 0.1 & 0.006 & -0.7\\
\cline{4-8}
 &  & \multirow{-2}{*}{\centering\arraybackslash $R \sim \bm{L}^{RY}$} & $N \sim \bm{L}^{A}$ & -0.001 & 0.1 & 0.004 & -0.5\\
\cline{3-8}
 &  &  & --- & -0.004 & 0.5 & 0.005 & -0.7\\
\cline{4-8}
 &  & \multirow{-2}{*}{\centering\arraybackslash $R \sim \bm{L}^{R}$} & $N \sim \bm{L}^{A}$ & -0.005 & 0.6 & 0.004 & -0.4\\
\cline{3-8}
 &  &  & --- & -0.004 & 0.5 & 0.006 & -0.7\\
\cline{4-8}
\multirow{-10}{*}{\centering\arraybackslash \shortstack{M-Bias\\w/ Mediator}} & \multirow{-10}{*}{\centering\arraybackslash -0.817} & \multirow{-2}{*}{\centering\arraybackslash $R \sim \bm{L}^{R} + A$} & $N \sim \bm{L}^{A}$ & -0.005 & 0.6 & 0.004 & -0.5\\
\hline
\end{tabular}\caption{Simulation results from hypothetical study \#1. Eligibility criteria: BMI $\geq$ 35 m/kg$^2$, A1c $\geq$ 5.7 \%, no previous initiation of bariatric surgery.}\label{table:results_1}
\end{table}

\begin{table}[H]
\scriptsize
\centering
\begin{tabular}{|>{}Sc|Sc|Sc|Sc|Sc|Sc|c|c|>{}c|}
\hline
\multicolumn{2}{|c|}{\textbf{Setting}} & \multicolumn{3}{c|}{\textbf{IPW Models}} & \multicolumn{2}{c|}{\textbf{Mean} $\bm{ \hat \psi}_{PP}$} & \multicolumn{2}{Sc|}{\textbf{Median} $\bm{ \hat{ \psi}_{PP}}$} \\
\cline{1-2} \cline{3-5} \cline{6-7} \cline{8-9}
\textbf{Missingness} & $\bm\psi_{PP}$ & \textbf{Statification} & $\bm W_{mk}^{R}$ & $\bm W_{mkt}^{N}$ & \textbf{Bias} & \textbf{\% Bias} & \textbf{Bias} & \textbf{\% Bias}\\
\hline
 &  &  &  & --- & -0.602 & 185.3 & -0.601 & 185.0\\
\cline{5-9}
 &  &  & \multirow{-2}{*}{\centering\arraybackslash ---} & $N \sim \bm{L}^{A}$ & -0.601 & 185.0 & -0.600 & 184.4\\
\cline{4-9}
 &  &  &  & --- & -0.605 & 186.0 & -0.604 & 185.8\\
\cline{5-9}
 &  &  & \multirow{-2}{*}{\centering\arraybackslash $R \sim \bm{L}^{RA}$} & $N \sim \bm{L}^{A}$ & -0.612 & 188.3 & -0.611 & 188.0\\
\cline{4-9}
 &  &  &  & --- & -0.571 & 175.6 & -0.569 & 175.1\\
\cline{5-9}
 &  &  & \multirow{-2}{*}{\centering\arraybackslash $R \sim \bm{L}^{RY}$} & $N \sim \bm{L}^{A}$ & -0.570 & 175.3 & -0.567 & 174.4\\
\cline{4-9}
 &  &  &  & --- & -0.568 & 174.8 & -0.567 & 174.5\\
\cline{5-9}
 &  &  & \multirow{-2}{*}{\centering\arraybackslash $R \sim \bm{L}^{R}$} & $N \sim \bm{L}^{A}$ & -0.577 & 177.4 & -0.575 & 176.7\\
\cline{4-9}
 &  &  &  & --- & -0.568 & 174.8 & -0.567 & 174.4\\
\cline{5-9}
 &  & \multirow{-10}{*}{\centering\arraybackslash Unstratified} & \multirow{-2}{*}{\centering\arraybackslash $R \sim \bm{L}^{R} + A$} & $N \sim \bm{L}^{A}$ & -0.577 & 177.4 & -0.575 & 176.7\\
\cline{3-9}
 &  &  &  & --- & -0.602 & 185.3 & -0.603 & 185.3\\
\cline{5-9}
 &  &  & \multirow{-2}{*}{\centering\arraybackslash $R \sim \bm{L}^{RA} $} & $N \sim \bm{L}^{A}$ & -0.610 & 187.7 & -0.609 & 187.4\\
\cline{4-9}
 &  &  &  & --- & -0.006 & 1.8 & -0.005 & 1.6\\
\cline{5-9}
 &  &  & \multirow{-2}{*}{\centering\arraybackslash $R \sim \bm{L}^{RY} $} & $N \sim \bm{L}^{A}$ & -0.005 & 1.5 & -0.004 & 1.2\\
\cline{4-9}
 &  &  &  & --- & 0.001 & -0.2 & 0.002 & -0.7\\
\cline{5-9}
\multirow{-16}{*}{\centering\arraybackslash M-Bias} & \multirow{-16}{*}{\centering\arraybackslash -0.325} & \multirow{-6}{*}{\centering\arraybackslash Stratified by $A$} & \multirow{-2}{*}{\centering\arraybackslash $R \sim \bm{L}^{R} $} & $N \sim \bm{L}^{A}$ & -0.008 & 2.4 & -0.007 & 2.1\\
\cline{1-9}
 &  &  &  & --- & -0.108 & 35.3 & -0.108 & 35.2\\
\cline{5-9}
 &  &  & \multirow{-2}{*}{\centering\arraybackslash ---} & $N \sim \bm{L}^{A}$ & -0.108 & 35.3 & -0.108 & 35.3\\
\cline{4-9}
 &  &  &  & --- & -0.108 & 35.4 & -0.110 & 36.0\\
\cline{5-9}
 &  &  & \multirow{-2}{*}{\centering\arraybackslash $R \sim \bm{L}^{RA}$} & $N \sim \bm{L}^{A}$ & -0.109 & 35.4 & -0.110 & 35.9\\
\cline{4-9}
 &  &  &  & --- & -0.141 & 45.8 & -0.141 & 46.0\\
\cline{5-9}
 &  &  & \multirow{-2}{*}{\centering\arraybackslash $R \sim \bm{L}^{RY}$} & $N \sim \bm{L}^{A}$ & -0.140 & 45.8 & -0.143 & 46.5\\
\cline{4-9}
 &  &  &  & --- & -0.140 & 45.6 & -0.139 & 45.4\\
\cline{5-9}
 &  &  & \multirow{-2}{*}{\centering\arraybackslash $R \sim \bm{L}^{R}$} & $N \sim \bm{L}^{A}$ & -0.140 & 45.8 & -0.141 & 45.9\\
\cline{4-9}
 &  &  &  & --- & -0.140 & 45.6 & -0.139 & 45.4\\
\cline{5-9}
 &  & \multirow{-10}{*}{\centering\arraybackslash Unstratified} & \multirow{-2}{*}{\centering\arraybackslash $R \sim \bm{L}^{R} + A$} & $N \sim \bm{L}^{A}$ & -0.140 & 45.8 & -0.141 & 46.0\\
\cline{3-9}
 &  &  &  & --- & -0.108 & 35.4 & -0.108 & 35.4\\
\cline{5-9}
 &  &  & \multirow{-2}{*}{\centering\arraybackslash $R \sim \bm{L}^{RA} $} & $N \sim \bm{L}^{A}$ & -0.109 & 35.5 & -0.108 & 35.2\\
\cline{4-9}
 &  &  &  & --- & 0.000 & 0.0 & -0.002 & 0.7\\
\cline{5-9}
 &  &  & \multirow{-2}{*}{\centering\arraybackslash $R \sim \bm{L}^{RY} $} & $N \sim \bm{L}^{A}$ & 0.000 & 0.0 & -0.001 & 0.4\\
\cline{4-9}
 &  &  &  & --- & 0.001 & -0.2 & -0.001 & 0.2\\
\cline{5-9}
\multirow{-16}{*}{\centering\arraybackslash \shortstack{Treatment\\Effect\\Heterogeneity}} & \multirow{-16}{*}{\centering\arraybackslash -0.307} & \multirow{-6}{*}{\centering\arraybackslash Stratified by $A$} & \multirow{-2}{*}{\centering\arraybackslash $R \sim \bm{L}^{R} $} & $N \sim \bm{L}^{A}$ & 0.000 & 0.0 & -0.001 & 0.3\\
\cline{1-9}
 &  &  &  & --- & -0.018 & 2.3 & -0.018 & 2.2\\
\cline{5-9}
 &  &  & \multirow{-2}{*}{\centering\arraybackslash ---} & $N \sim \bm{L}^{A}$ & -0.016 & 2.0 & -0.017 & 2.0\\
\cline{4-9}
 &  &  &  & --- & -0.020 & 2.5 & -0.019 & 2.4\\
\cline{5-9}
 &  &  & \multirow{-2}{*}{\centering\arraybackslash $R \sim \bm{L}^{RA}$} & $N \sim \bm{L}^{A}$ & -0.018 & 2.3 & -0.018 & 2.2\\
\cline{4-9}
 &  &  &  & --- & -0.075 & 9.2 & -0.074 & 9.1\\
\cline{5-9}
 &  &  & \multirow{-2}{*}{\centering\arraybackslash $R \sim \bm{L}^{RY}$} & $N \sim \bm{L}^{A}$ & -0.073 & 9.0 & -0.072 & 8.9\\
\cline{4-9}
 &  &  &  & --- & -0.078 & 9.6 & -0.078 & 9.6\\
\cline{5-9}
 &  &  & \multirow{-2}{*}{\centering\arraybackslash $R \sim \bm{L}^{R}$} & $N \sim \bm{L}^{A}$ & -0.076 & 9.4 & -0.076 & 9.4\\
\cline{4-9}
 &  &  &  & --- & -0.078 & 9.6 & -0.078 & 9.6\\
\cline{5-9}
 &  & \multirow{-10}{*}{\centering\arraybackslash Unstratified} & \multirow{-2}{*}{\centering\arraybackslash $R \sim \bm{L}^{R} + A$} & $N \sim \bm{L}^{A}$ & -0.076 & 9.4 & -0.076 & 9.4\\
\cline{3-9}
 &  &  &  & --- & -0.022 & 2.8 & -0.020 & 2.5\\
\cline{5-9}
 &  &  & \multirow{-2}{*}{\centering\arraybackslash $R \sim \bm{L}^{RA} $} & $N \sim \bm{L}^{A}$ & -0.020 & 2.5 & -0.019 & 2.3\\
\cline{4-9}
 &  &  &  & --- & 0.001 & -0.2 & 0.003 & -0.4\\
\cline{5-9}
 &  &  & \multirow{-2}{*}{\centering\arraybackslash $R \sim \bm{L}^{RY} $} & $N \sim \bm{L}^{A}$ & 0.003 & -0.4 & 0.005 & -0.6\\
\cline{4-9}
 &  &  &  & --- & -0.002 & 0.2 & -0.001 & 0.1\\
\cline{5-9}
\multirow{-16}{*}{\centering\arraybackslash \shortstack{M-Bias\\w/ Mediator}} & \multirow{-16}{*}{\centering\arraybackslash -0.811} & \multirow{-6}{*}{\centering\arraybackslash Stratified by $A$} & \multirow{-2}{*}{\centering\arraybackslash $R \sim \bm{L}^{R} $} & $N \sim \bm{L}^{A}$ & 0.000 & 0.0 & 0.002 & -0.2\\
\hline
\end{tabular}\caption{Simulation results from hypothetical study \#2. Eligibility criteria: BMI $\geq$ 35 m/kg$^2$, no previous initiation of bariatric surgery.}\label{table:results_2}
\end{table}

\begin{table}[H]
\centering
\begin{tabular}{|Sc|Sc|Sc|}
\hline
\textbf{Outcome} & \textbf{Intention-To-Treat} ($\bm{e^{\hat{\psi}_{ITT}}}$) & \textbf{Per-Protocol} ($\bm{e^{\hat{\psi}_{PP}}}$)\\
\hline
Any Microvascular Event & 0.438 (0.382, 0.497) & 0.429 (0.375, 0.489)\\
\hline
Nephropathy & 0.340 (0.239, 0.458) & 0.331 (0.236, 0.452)\\
\hline
Neuropathy & 0.436 (0.367, 0.521) & 0.428 (0.360, 0.512)\\
\hline
Retinopathy & 0.507 (0.400, 0.608) & 0.491 (0.394, 0.600)\\
\hline
\end{tabular}
\caption{Discrete hazard ratio estimates and 95\% confidence intervals for the effect of bariatric surgery on microvascular outcomes. Confidence intervals were computed utilizing 1,000 bootstrap replications at the subject ($k$) level. Eligibility ($E_{mk}$) was assessed utilizing a 12 month lookback window  prior to the start of trial $m$ for blood glucose measures and a 3 month lookback window for BMI measures.}\label{table:data_app}
\end{table}

\end{document}


\title{Supplementary Materials for ``Adjusting for Selection Bias Due to Missing Eligibility Criteria in Emulated Target Trials''}
\author[1]{Luke Benz}
\author[1]{Rajarshi Mukherjee}
\author[1,2,3]{Rui Wang}
\author[4]{David Arterburn}
\author[5]{Heidi Fischer}
\author[6]{Catherine Lee}
\author[7,8]{Susan M. Shortreed}
\author[1]{Sebastien Haneuse}
\affil[1]{Department of Biostatistics,
Harvard T.H. Chan School of Public Health, Boston, MA, USA}
\affil[2]{Department of Population Medicine, Harvard Pilgrim Health Care Institute, Boston, MA, USA}
\affil[3]{Department of Population Medicine, Harvard Medical School, Boston, MA, USA}
\affil[4]{Kaiser Permanente Washington Health Research Institute, Seattle, WA, USA}
\affil[5]{Department of Research \& Evaluation, Kaiser Permanente Southern California, Pasadena, CA, USA}
\affil[6]{Department of Epidemiology and Biostatistics, University of California San Francisco, San Francisco, CA, USA}
\affil[7]{Biostatistics Division, Kaiser Permanente Washington Health Research Institute, Seattle, WA, USA}
\affil[8]{Department of Biostatistics, University of Washington School of Public Health, Seattle, WA, USA}

\date{
    \today 
}

\maketitle


\makeatletter
\renewcommand \thesection{S\@arabic\c@section}
\renewcommand\thetable{S\@arabic\c@table}
\renewcommand \thefigure{S\@arabic\c@figure}
\renewcommand \theequation{S\@arabic\c@equation}
\makeatother
\setcounter{figure}{0}
\setcounter{table}{0}
\setcounter{equation}{0}

\tableofcontents

\newpage


\section{Assumptions}\label{sec:supp_assumptions}

Fitting the model in Equation 3 of the main paper estimates the parameter $\tilde \theta$. Without care, $\tilde \theta \neq \psi$, the parameter of interest, due to the possibility of selection bias ($\theta \neq \psi$) as well as non-exchangability of treated and untreated subjects due to the possibility of confounding, differential non-adherence, and differential censoring ($\tilde \theta \neq \theta$). Recovery of $\psi$ by fitting the model in Equation 3 with inverse probability weights $W_{mkt} = W^{A}_{mk} \times W^{C}_{mkt} \times W^{N}_{mkt} \times W^{R}_{mk}$ requires certain assumptions \citep{robins2000marginal, hernan2004structural, hernan2024, dickerman2019, danaei2013observational, cain2010, karim2018comparison}. One such assumption is the missing at random (MAR) assumption outlined in Equation 7. Additional assumptions ensure component weights address other potential sources of bias, and guarantee weights are well defined. The complete list of assumptions, as well as some additional commentary, is detailed below.

\begin{enumerate}
    \item \textbf{Consistency of outcomes} 
    $$Y^{(\bar a)}_{mkt} = Y_{mkt}~|~\bar A_{mkt} = \bar a, E_{mk} = 1$$
    \item \textbf{No unmeasured confounding} 
    $$Y_{mkt}^{(\bar a)} \indep A_{mkt} ~|~ \bm L_{mk}, E_{mk} = 1$$
    \item \textbf{Positivity of treatment} 
    $$\epsilon < P(A_{mk} = 1 ~|~ \bm L_{mk}, E_{mk}= 1) < 1 - \epsilon~\text{for some}~\epsilon > 0$$
    \item \textbf{Positivity of remaining uncensored} 
    $$P(C_{mkt} = 0~|~\bar C_{mk(t-1)} = 0, \bar N_{mk(t-1)} = 0, \bm L_{mkt}, E_{mk} = 1, A_{mk}) > \epsilon > 0$$
    \item \textbf{Positivity of remaining adherent to baseline treatment}
    $$P(N_{mkt} = 0~|~\bar C_{mk(t-1)} = 0, \bar N_{mk(t-1)} = 0, \bm L_{mkt}, E_{mk} = 1, A_{mk}) >  \epsilon > 0$$
    \item  \textbf{Conditional independence of censoring and non-adherence}
    $$N_{mkt} \indep C_{mkt} ~|~ \bar{N}_{mk(t-1)} = 0, \bar{C}_{mk(t-1)} = 0, \bm L_{mkt}, E_{mk} = 1, A_{mk}$$ 
    \item \textbf{Eligibility status missing at random}
    $$R_{mk} \indep E_{mk}~|~\bar{\bm L}^c_{mk}, \bar{A}_{mk}, C_{mk} = 0$$
    \item \textbf{Positivity of observing eligibility status} 
    $$P(R_{mk}~|~ \bar{\bm L}^c_{mk}, \bar{A}_{mk}, C_{mk} = 0) > \epsilon > 0$$
\end{enumerate}

Assumptions 1-3 are standard causal assumptions among eligible patients for addressing confounding via inverse probability weighting (IPW). Assumptions 4-5 are positivity assumptions that ensure inverse probability weights to address non-adherence and censoring are well defined. Assumption 6 facilitates separation of weights to address differential non-adherence and differential censoring with distinct inverse probability weights. 
We explicitly distinguish censoring due to non-adherence, which is sometimes referred to as artificial censoring \citep{emilsson2018, maringe2020reflection, danaei2013observational, cain2010, hajage2022extracorporeal,karim2018comparison, gran2010sequential, petito2020, keogh2023causal}, from censoring due to loss to follow up, development of a competing risk (e.g. death), or other reasons. Artificial censoring due to non-adherence is a direct by-product of underlying treatment mechanisms, making it fundamentally different from other forms of censoring, which are due to distinct mechanisms. While one might choose not to worry about either non-adherence (if intention-to-treat effects are of interest) or canonical censoring (if differential loss to follow up is not a concern), keeping these processes distinct in our notational framework allows for maximal flexibility when addressing these issues. Notably, the two events $I(N_{mkt} = 0)$ and $I(C_{mkt} = 0)$ may depend on different variables within $\bm L_{mkt}$, or on the same variables to different extents. Modeling weights for each event separately makes it more flexible than modeling the joint even $I(C_{mkt} = 0, N_{mkt} = 0)$. 

Violations of Assumption 6 seem rare, as violations of such an assumption involve someone being lost to follow-up imminently before non-adhering to their baseline treatment status, or vice versa. In the context of our running bariatric surgery example, such a violation could involve someone disappearing from the electronic health record (EHR) system to receive bariatric surgery at a separate healthcare facility. Under such an an example, the subject would be lost to follow-up at the same time they became non-adherent to their previous non-surgical treatment status. Should such an assumption be exceedingly unrealistic in other settings, analysts may choose not to make this assumption and instead proceed by joint modeling of $I(C_{mkt} = 0, N_{mkt} = 0)$.

Assumption 7 is the novel missing at random assumption which allows us to address selection bias via IPW, with associated positivity assumption (Assumption 8) ensuring weights are well defined.

Finally, implicit in our framework is the assumption that there is no missing data in other components of the problem, namely in $\bm L^c, A$ and $Y$. As is common practice when working with EHR data, we assume absence of a procedure or diagnostic code implies the absence of a condition, which is relevant for establishing both bariatric surgery ($A$) as well as component microvascular outcomes of interest ($Y$) in our application.

\section{Simulations}\label{sec:supp_simulations}

\subsection{EHR-Based Simulation Infrastructure}
\label{sec:simulation_DGP}
First, covariate values $\bm{L}_{k0}$ for BMI, A1c, smoking status, comorbidity score, gender, race, and health care site were sampled from observed non-surgical patients in the DURABLE database. Critically, this preserved the joint distribution of these variables from the electronic health records, and thus their complex correlation structure. Next, time-varying BMI and A1c trajectories (and through them, time-varying eligibility status) were simulated in a manner resembling post-surgical patient outcomes \cite{arterburn2014bariatric, courcoulas2013weight}.

Trajectories in the absence of surgery were simulated as follows:

\begin{equation}\label{eq:presurg_traj}
\begin{aligned}
    \text{BMI}_{kt} &= \text{BMI}_{k0}\biggr[1 + (\bm\beta^T_1\bm{L}_{k0} + \gamma_{1k})t + q_{k}(\bm\beta^T_2\bm{L}_{k0} + \gamma_{2k})t^2 + \epsilon^{\text{(BMI)}}_{kt}\biggr] \\
    \text{A1c}_{kt} &= \text{A1c}_{k0}\biggr[1 + (\bm\beta^T_3\bm{L}_{k0} + \gamma_{3k})t + \epsilon^{(\text{AIc})}_{kt}\biggr] 
\end{aligned}
\end{equation}

\noindent where $\bm \beta_1, \bm \beta_2, \bm \beta_3$ denote fixed effects, $\gamma_{1k} \sim N(0, \tau_{1}^2), \gamma_{2k} \sim N(0, \tau_{2}^2), \gamma_{3k} \sim N(0, \tau_{3}^2)$ denote subject specific random, $q_k$ is a binary indicator of a quadratic BMI trajectory generated from the logistic regression model $\logit[P(q_{k} = 1~|~\bm L_{k0})] = \delta_0 + \bm \delta_1^T \bm L_{k0}$, and finally $\epsilon^{\text{(BMI)}}_{kt} \sim N(0, \sigma_{\text{BMI}}^2)$ and $\epsilon^{\text{(A1c)}}_{kt} \sim N(0, \sigma^2(\text{A1c}_{k0}))$ are random errors. Of particular note, random errors in the generation of BMI trajectories are homoskedastic, while random errors in generation of A1c trajectories are heteroskedastic with variances that are a function of initial A1c values (A1c$_{k0}$). The specification in Equation \ref{eq:presurg_traj} gives us control over the trajectories on the basis of modeling relative weight/A1c change as in previous studies \citep{arterburn2020, li2021fiveyear, courcoulas2013weight}, with a simple transformation recovering values on the absolute scale.

Next, at each timepoint $t$, treatment assignments to undergo bariatric surgery were simulated from a logistic regression model, $\logit[P(A_{kt} = 1~|~ \bar A_{k(t-1)} = 0, \bm L_{k0}, \bm L_{kt}, \text{BMI}_{kt} \geq 35\text{kg/m}^2)] = \alpha_{0} + \bm \alpha_1^T \bm L_{k0} + \bm \alpha_2^T \bm L_{kt}$, where $\bm L_{kt}$ denotes the current values of any time-varying covariates such as BMI and A1c, and $\text{BMI}_{kt} \geq 35\text{kg/m}^2$ indicates that a subject met the treatment eligibility criteria to undergo bariatric surgery \citep{kaiserpermanente2023}. An analogous model was used to determine bariatric surgery type, Roux-en-Y gastric bypass (RYGB) or vertical sleeve gastrectomy (VSG), among those sampled to undergo treatment. Namely, letting $A'_{kt} = 1$ denote RYGB, bariatric surgery type was assigned via the logsitic regression model $\logit[P(A'_{kt} = 1~|~A_{kt} = 1, \bar A_{k(t-1)} = 0, \bm L_{k0}, \bm L_{kt})] = \pi_{0} + \bm \pi_1^T \bm L_{k0} + \bm \pi_2^T \bm L_{kt}$.

\begin{figure}[ht]
    \centering
    \includegraphics[width=\textwidth]{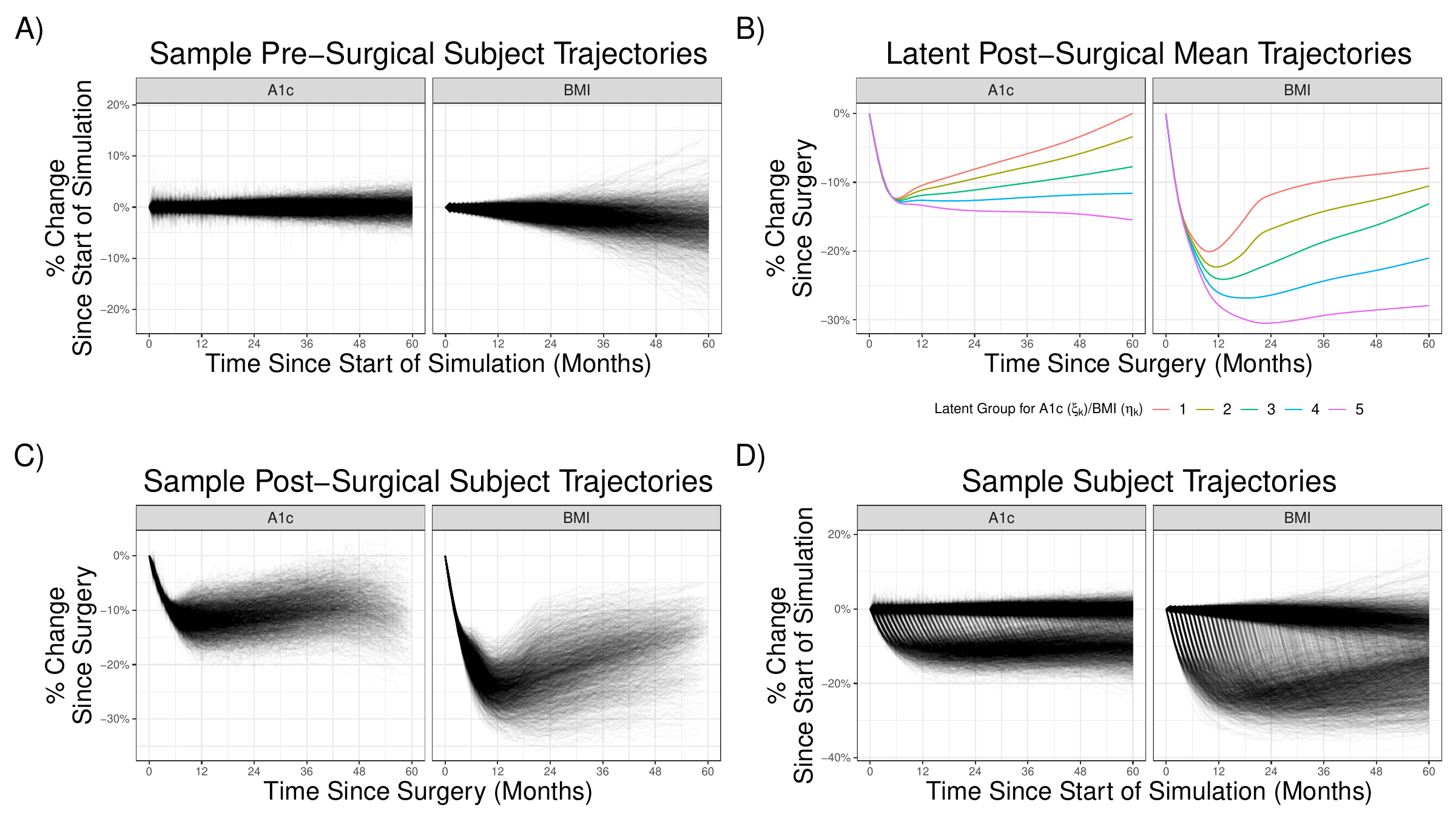}
    \caption{Summary of BMI/A1c trajectory generation. Note that for ease of display, trajectories are shown on a relative scale. In simulations however, they exist on an absolute scale. \textbf{(A)} A1c and BMI are first generated in the absence of surgery. \textbf{(B)} Upon assignment to bariatric surgery, subjects are randomized into a latent BMI/A1c class which determines post-surgical mean trajectories. \textbf{(C)} Next, trajectory generation continues post-surgery. \textbf{(D)} Finally, pre- and post-surgical trajectories are combined.
    }
    \label{fig:simulation_map}
\end{figure}

Upon receiving bariatric surgery, the remainder of a subject's BMI and A1c trajectories through the conclusion of follow up (at time $t = T_{\text{max}}$) were generated from a different set of models than non-surgical trajectories. To be specific, let $\nu_k$ denote treatment time for subject $k$ undergoing surgery, and let $\eta_k, \xi_k \in \{1, 2, 3, 4, 5\}$ denote latent post-surgical trajectories for BMI and A1c, respectively. Five latent classes were chosen for post-surgical trajectories to match existing literature on heterogeneity of long term weight loss outcomes following bariatric surgery \citep{courcoulas2013weight}. In our simulations, latent post-surgical classes were sampled for subjects undergoing bariatric surgery from the ordinal logistic regression models:

\begin{equation}\label{eq:latent_classes}
\begin{aligned}
\logit[P(\eta_k \leq j)~|~A_{k\nu_{k}} = 1, A'_{k\nu_{k}}, \bm L_{k0}, \bm L_{k1})] &= \lambda_{0j} + \bm \lambda_1^{T} \bm L_{k0} + \bm \lambda_2^T \bm L_{k\nu_k} + \lambda_3 A'_{k\nu_{k}} \\
\logit[P(\xi_k \leq j)~|~A_{k\nu_{k}} = 1, A'_{k\nu_{k}}, \bm L_{k0}, \bm L_{k1})] &= \phi_{0j} + \bm \phi_1^{T} \bm L_{k0} + \bm \phi_2^T \bm L_{k\nu_k} + \phi_3 A'_{k\nu_{k}}
\end{aligned}
\end{equation}

\noindent For time $t \in (\nu_k, T_{\text{max}}]$, post-surgical trajectories were generated as follows:

\begin{equation}\label{eq:postsurg_traj}
\begin{aligned}
    \text{BMI}_{kt} &= \text{BMI}_{k\nu_k}\biggr[1 + \sum_{j = 1}^5 I(\eta_k = j)b(t - \nu_k)^T(\bm \beta^{\text{(BMI)}}_j + \gamma_{4k}) +  \epsilon^{\text{(BMI)}}_{kt}\biggr] \\
    \text{A1c}_{kt} &= \text{A1c}_{k\nu_k}\biggr[1 +\sum_{j = 1}^5 I(\xi_k = j)b(t - \nu_k)^T(\bm \beta^{\text{(A1c)}}_j + \gamma_{5k}) +  \epsilon^{\text{(A1c)}}_{kt}\biggr] \\
\end{aligned}
\end{equation}

\noindent where $b(t)^T$ denotes a natural cubic spline on time, with coefficients $\bm \beta_j^{(BMI)}$ and $\bm \beta_j^{(A1c)}$ corresponding to latent class $j$. Specific values for $\bm \beta_j^{(BMI)}$ and $\bm \beta_j^{(A1c)}$ were choosen based on a combination of empirical fits in the DURABLE database and prior work on long term outcomes following bariatric surgery, in particular work by Arterburn et al. \cite{arterburn2020} and Courcoulas and colleagues \cite{courcoulas2013weight}. In Equation \ref{eq:postsurg_traj}, $\gamma_{4k} \sim N(0, \tau_4^2)$ and $\gamma_{5k} \sim N(0, \tau_5^2)$ denote subject-specific random effects. A graphical overview of BMI and A1c trajectory generation is outlined in Figure \ref{fig:simulation_map}.

Outcomes, in the form of monthly binary indicators of an event were generated from the logistic regression model $\logit[P(Y_{kt} = 1 ~|~ A_{kt}, \bm L_{k0}, \bm L_{kt})] = \omega_{0} + \bm \omega_1^T \bm L_{k0} + \bm \omega_2^T \bm L_{kt}$ for $t \in [0, \min(T^*_{k}, T_\text{max})]$. Finally, complete case indicators of eligibility ascertainment were sampled from the logistic regression model $\logit[P(R_{kt} = 1 ~|~ A_{kt}, \bm L_{k0}, \bm L_{kt})] = \rho_{0} + \bm \rho_1^T \bm L_{k0} + \bm \rho_2^T \bm L_{kt}$. In time periods where $R_{kt} = 0$ values of $\text{BMI}_{kt}$ and $\text{A1c}_{kt}$ were set to be missing.

\begin{figure}[ht]
    \centering
    \includegraphics[width=0.7\textwidth]{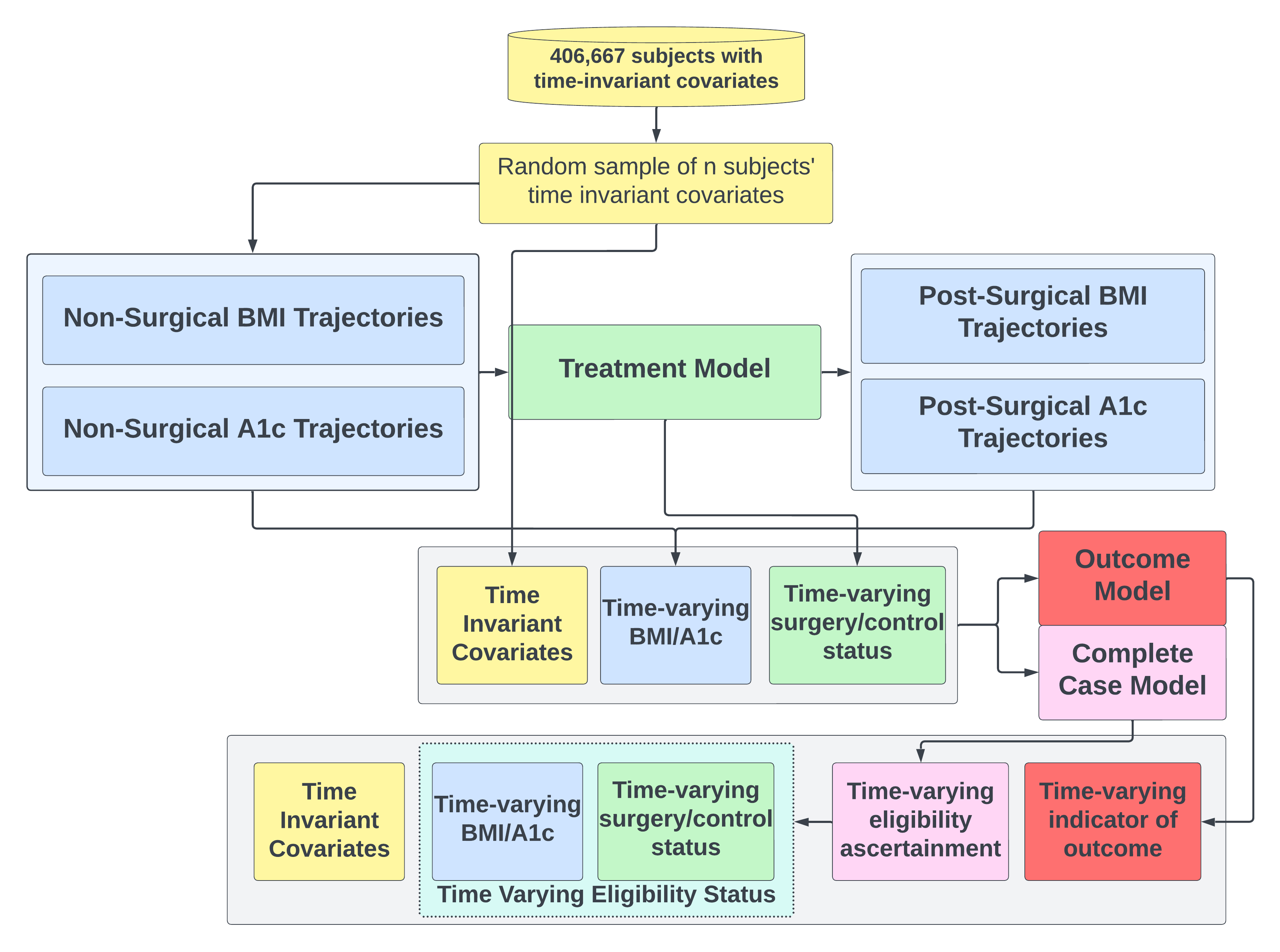}
    \caption{Overview of simulation infrastructure. First, a set of time-invariant covariates are sampled from observed non-surgical patients in the DURABLE database. Next, BMI and A1c trajectories are generated in the absence of surgery, until either the end of study or assignment to undergo surgery, at which point post-surgical trajectories are sampled (Figure \ref{fig:simulation_map}). Finally, outcomes and eligibility ascertainment indicators are sampled for each time, and eligible status ($E_{kt}$) and associated eligibility defining covariates $\bm L_{kt}^e$ are set to be missing, at which point application of analytical methods proceeds.}
    \label{fig:simulation_infrastructure}
\end{figure}

While this simulation infrastructure has many moving pieces and seems highly complex, such complexity helps adequately characterize and understand the problem of missing eligibility criteria in the context of EHR-based observational studies. Figure \ref{fig:simulation_infrastructure} provides an overview of the EHR-based simulation infrastructure developed as part of this work. Specific covariates used in each model, as well as exact coefficient values are outlined in Section \ref{sec:supp_params}.

\subsection{Hypothetical Study Designs}\label{sec:supp_designs}
In the first hypothetical study, eligibility criteria for trial $m$ required a BMI measurement exceeding 35 kg/m$^2$ and an A1c measurement of at least $5.7\%$ within the first time period of the trial, and that subjects had not previously undergone bariatric surgery, thereby restricting the study population to those who were simultaneously pre-diabetic and eligible for consideration to undergo bariatric surgery at the start of the trial. Under this scenario, eligibility ascertainment is a problem equally affecting both study arms. Undergoing surgery implies a patient must have met the clinical recommendation for BMI of at least 35 kg/m$^2$ at the time of surgery, and hence we are able to conclude that they meet the study BMI criteria, even if no BMI measurement was available at the time of surgery. However, due to the additional A1c threshold in the eligibility criteria, knowing a patient underwent surgery and met the specific BMI eligibility criterion says nothing about their A1c, and thus their study eligibility can not be deduced at that point in time.  

In the second hypothetical study, the only eligibility criterion consisted of having a BMI above 35 kg/m$^2$ during the initial time period of trial $m$, and not having previously undergone bariatric surgery. By the argument in the previous paragraph, the inability to ascertain eligibility is a one sided problem affecting only non-surgical patients. Note that the implication that undergoing treatment makes one eligible for an observational study will only be true when the requisite criteria for treatment initiation (should any exist) are identical to the eligibility criteria of the corresponding observational study. For example, had a emulated trial been interested in a different population of more severely obese patients, and required a BMI of 40 kg/m$^2$, a cutoff different than the clinical minimum of 35 kg/m$^2$ in the Kaiser system, treatment initiation would not directly imply eligibility for the corresponding observational study. 

In practice, it is often the case that eligibility for initiation of a treatment may be the same as, or at least a subset of eligibility for an observational study, especially if the observation is comparing the effectiveness of treatment vs. no treatment. In such settings, in order to make rigorous comparisons between initiators and non-initiators, analysis is restricted to those who underwent initiation of a treatment and those who were hypothetically eligible for initiation but did not undergo treatment.

In both hypothetical studies considered, $\bm L^{RA}=(\text{Smoking Status}, \text{Health Care Site})$ and $\bm L^{RY} = (\text{Comorbidity Score})$ when missing data were generated the mechanisms outlined in Figures 3a and 3b. When missing data were generated in a manner outlined in Figure 3c,  $\bm L^{RA}=(\text{Health Care Site})$ and $\bm L^{RY} = (\text{Comorbidity Score, Insulin Usage})$. In the context of these simulations, mediators $\bm M$ are BMI and A1c values, whose respective observation frequency vary in practice. In all cases, smoking status was the only covariate which influenced the probability to undergo bariatric surgery $(\bm L^{A})$ with health care site influencing the type of bariatric surgery a subject underwent.

\subsection{Analytical Approaches}
\label{sec:supp_simulation_approaches}
In total, 1000 simulated datasets were generated for each combination  of study question and missing data mechanism. Each dataset consisted of $K = 10,000$ subjects followed over a period of $T_\text{max} = 36$ months. Data were analyzed using a sequential target trial emulation framework with trials in each of the first 12 months. $M = 12$ was chosen to keep the final analysis dataset at a reasonable size. M

In the extreme where all $K$ subjects are eligible for all $M$ trials, and there isn't censoring/non-adherence/incidence of the outcome, which would remove certain observations from the final analysis dataset, the size of the pooled analysis data set is $T_{\text{max}}$ is $K\sum_{m = 1}^{M}(T_\text{max} - m + 1) = KM(T_{\text{max}} - \frac{M+1}{2})$ observations. While of course eligibility/censoring/non-adherence/incidence of the outcome will all reduce the size of the final analysis dataset, choosing $M = 12$ instead of the theoretical maximum of $M = T_\text{max}$ reduces the size of a single simulated dataset by tens or even hundreds of thousands of observations. More importantly, however, choosing $M < T_{\text{max}}$ is necessary to allow sufficient follow up time for subjects in later trials. In our data application, the last trial was started such that nearly 5 years of follow-up were possible. 

Under our simulations, we considered bariatric surgery as a non-reversible treatment, and thus someone initiating treatment is by definition always adherent to their baseline treatment status during the trial in which they enter as a surgical initiator. As such, non-adherence weights were only estimated for subjects entering each trial $m$ as non-initiators of surgery, that is eligible subjects for whom $A_{mk} = 0$, with weights $W_{mkt}^N = 1$ for all initiators of surgery. Fitting separate models for non-adherence stratified by treatment status has been considered in existing target trial emulations \citep{dickerman2019, massol2023levosimendan, emilsson2018, danaei2013observational, caniglia2019}, especially in the context of one sided non-adherence \citep{hajage2022extracorporeal, maringe2020reflection}.

True values of $\psi_{PP}$ in each setting were computed by fitting Equation (6) before missing eligibility criteria was imposed, with inverse probability weights for non-adherence from correctly specified models, and averaging the results across the 1000 simulated datasets. The true value of $\psi_{PP}$ was computed in this manner due to the complex nature through which outcomes were generated, which involved a series of high dimensional, non-linear, conditional models. As such, no closed form for the marginal parameter of interest was readily attainable.

\subsection{Simulation Parameters}\label{sec:supp_params}
Spline coefficients for post-surgical BMI and A1c trajectories, as displayed in Figure 4, are shown in Table \ref{table:spline_coef}. Coefficient values used in simulation studies are available in Table \ref{table:sim_coef}.

\begin{table}[H]
\scriptsize
\centering
\begin{tabular}{|>{}Sc|Sc|Sc|c|c|>{}c|}
\hline
\textbf{Measure} & $\bm\beta_1$ & $\bm\beta_2$ & $\bm\beta_3$ & $\bm\beta_4$ & $\bm\beta_5$\\
\hline
 & -0.181 & -0.186 & -0.191 & -0.196 & -0.201\\
\cline{2-6}
 & -0.205 & -0.221 & -0.232 & -0.245 & -0.256\\
\cline{2-6}
 & -0.198 & -0.225 & -0.243 & -0.263 & -0.281\\
\cline{2-6}
 & -0.178 & -0.216 & -0.240 & -0.268 & -0.292\\
\cline{2-6}
 & -0.154 & -0.204 & -0.234 & -0.269 & -0.300\\
\cline{2-6}
 & -0.125 & -0.175 & -0.225 & -0.268 & -0.305\\
\cline{2-6}
 & -0.111 & -0.161 & -0.211 & -0.261 & -0.305\\
\cline{2-6}
 & -0.096 & -0.140 & -0.184 & -0.241 & -0.292\\
\cline{2-6}
 & -0.047 & -0.085 & -0.122 & -0.187 & -0.244\\
\cline{2-6}
 & -0.207 & -0.238 & -0.269 & -0.342 & -0.405\\
\cline{2-6}
\multirow{-11}{*}{\centering\arraybackslash BMI} & 0.003 & -0.022 & -0.047 & -0.127 & -0.197\\
\cline{1-6}
 & -0.110 & -0.115 & -0.120 & -0.125 & -0.130\\
\cline{2-6}
 & -0.093 & -0.104 & -0.117 & -0.129 & -0.141\\
\cline{2-6}
 & -0.067 & -0.084 & -0.105 & -0.124 & -0.143\\
\cline{2-6}
 & -0.008 & -0.031 & -0.060 & -0.086 & -0.112\\
\cline{2-6}
 & -0.130 & -0.159 & -0.196 & -0.229 & -0.262\\
\cline{2-6}
\multirow{-6}{*}{\centering\arraybackslash A1c} & 0.024 & -0.011 & -0.056 & -0.096 & -0.136\\
\hline
\end{tabular}\caption{Spline coefficients $\bm\beta_j$ for post-surgical BMI and A1c trajectories. A subject's set of coefficients are determined by latent post-surgical classes $\eta_k$ (for BMI) and $\xi_{k}$ (for A1c). Knots for BMI splines are at $t = 3, 6, 9, 12, 15, 18, 21, 24, 36$, and $48$ months, while knots for A1c splines are at $3, 9, 15, 30$ and $48$ months.}\label{table:spline_coef}
\end{table}

\begin{table}[H]
\tiny
\centering
\begin{tabular}{|>{}Sc|Sc|Sc|c|c|>{}c|}
\hline
\multicolumn{3}{|Sc|}{\textbf{Model Information}} & \multicolumn{3}{Sc|}{\textbf{Missingness Mechanism}} \\
\cline{1-3} \cline{4-6}
\textbf{Component Model} & \textbf{Coefficient} & \textbf{Variable} & \textbf{Fig. 3a} & \textbf{Fig. 3b} & \textbf{Fig. 3c}\\
\hline
 &  & \texttt{(Intercept)} & $-5 \times 10^{-4}$ & $-5 \times 10^{-4}$ & $-5 \times 10^{-4}$\\
\cline{3-6}
 &  & \texttt{gender} & 0 & 0 & $-1 \times 10^{-4}$\\
\cline{3-6}
 &  & \texttt{race} & 0 & 0 & $-2 \times 10^{-4}$\\
\cline{3-6}
 & \multirow{-4}{*}{\centering\arraybackslash $\bm \beta_1$} & \texttt{insulin} & 0 & 0 & $5 \times 10^{-4}$\\
\cline{2-6}
 &  & \texttt{(Intercept)} & 0 & 0 & 0\\
\cline{3-6}
 & \multirow{-2}{*}{\centering\arraybackslash $\bm \beta_2$} & \texttt{insulin} & 0 & 0 & $1 \times 10^{-5}$\\
\cline{2-6}
 &  & \texttt{(Intercept)} & -1.10 & -1.10 & -1.10\\
\cline{3-6}
 & \multirow{-2}{*}{\centering\arraybackslash $\bm \delta$} & \texttt{insulin} & 0 & 0 & 0.50\\
\cline{2-6}
 & $\sigma^2_{\text{BMI}}$ & \texttt{---} & $3.2 \times 10^{-3}$ & $3.2 \times 10^{-3}$ & $3.2 \times 10^{-3}$\\
\cline{2-6}
 & $\tau^2_1$ & \texttt{---} & $5 \times 10^{-4}$ & $5 \times 10^{-4}$ & $5 \times 10^{-4}$\\
\cline{2-6}
\multirow{-11}{*}{\centering\arraybackslash Pre-Surgery BMI} & $\tau^2_2$ & \texttt{---} & $2 \times 10^{-5}$ & $2 \times 10^{-5}$ & $2 \times 10^{-5}$\\
\cline{1-6}
 &  & \texttt{(Intercept)} & 0 & 0 & 0\\
\cline{3-6}
 & \multirow{-2}{*}{\centering\arraybackslash $\bm \beta_3$} & \texttt{insulin} & 0 & 0 & $-1 \times 10^{-3}$\\
\cline{2-6}
 & $\sigma^2_{\text{A1c}}$ & \texttt{I(hgba1c\string^2)} & $1 \times 10^{-3}$ & $1 \times 10^{-3}$ & $1 \times 10^{-3}$\\
\cline{2-6}
\multirow{-4}{*}{\centering\arraybackslash Pre-Surgery A1c} & $\tau^3_2$ & \texttt{---} & $3 \times 10^{-4}$ & $3 \times 10^{-4}$ & $3 \times 10^{-4}$\\
\cline{1-6}
 &  & \texttt{(Intercept)} & -3.18 & -3.18 & -3.18\\
\cline{3-6}
 &  & \texttt{smoking\_status[current]} & -2.00 & -2.00 & -2.00\\
\cline{3-6}
 & \multirow{-3}{*}{\centering\arraybackslash $\bm\alpha$} & \texttt{smoking\_status[former]} & -0.75 & -0.75 & -0.75\\
\cline{2-6}
 &  & \texttt{(Intercept)} & 0.41 & 0.41 & 0.41\\
\cline{3-6}
\multirow{-5}{*}{\centering\arraybackslash Treatment} & \multirow{-2}{*}{\centering\arraybackslash $\bm\pi$} & \texttt{site} & 0.69 & 0.69 & 0.69\\
\cline{1-6}
 & $\lambda_{01}$ & \texttt{(Intercept)} & -2.94 & -2.94 & -2.94\\
\cline{2-6}
 & $\lambda_{02}$ & \texttt{(Intercept)} & -1.10 & -1.10 & -1.10\\
\cline{2-6}
 & $\lambda_{03}$ & \texttt{(Intercept)} & 1.10 & 1.10 & 1.10\\
\cline{2-6}
 & $\lambda_{04}$ & \texttt{(Intercept)} & 2.94 & 2.94 & 2.94\\
\cline{2-6}
 &  & \texttt{elix\_score} & 0 & 0 & -5.00\\
\cline{3-6}
 &  & \texttt{bs\_type[rygb]} & 0 & 0 & 4.00\\
\cline{3-6}
 & \multirow{-3}{*}{\centering\arraybackslash $\bm\lambda_1$} & \texttt{insulin} & 0 & 0 & -3.00\\
\cline{2-6}
\multirow{-8}{*}{\centering\arraybackslash Post-Surgery BMI} & $\tau^2_4$ & \texttt{---} & $2.5 \times 10^{-2}$ & $2.5 \times 10^{-2}$ & $2.5 \times 10^{-2}$\\
\cline{1-6}
 & $\phi_{01}$ & \texttt{(Intercept)} & -2.94 & -2.94 & -1.39\\
\cline{2-6}
 & $\phi_{02}$ & \texttt{(Intercept)} & -1.10 & -1.10 & -0.20\\
\cline{2-6}
 & $\phi_{03}$ & \texttt{(Intercept)} & 1.10 & 1.10 & 1.73\\
\cline{2-6}
 & $\phi_{04}$ & \texttt{(Intercept)} & 2.94 & 2.94 & 2.94\\
\cline{2-6}
 &  & \texttt{elix\_score} & 0 & 0 & -5.00\\
\cline{3-6}
 &  & \texttt{bs\_type[rygb]} & 0 & 0 & 4.00\\
\cline{3-6}
 & \multirow{-3}{*}{\centering\arraybackslash $\bm\phi_1$} & \texttt{insulin} & 0 & 0 & -3.00\\
\cline{2-6}
\multirow{-8}{*}{\centering\arraybackslash Post-Surgery A1c} & $\tau^2_5$ & \texttt{---} & $2.5 \times 10^{-2}$ & $2.5 \times 10^{-2}$ & $2.5 \times 10^{-2}$\\
\cline{1-6}
 &  & \texttt{(Intercept)} & -4.18 & -3.66 & -9.21\\
\cline{3-6}
 &  & \texttt{elix\_score} & 1.20 & 0.25 & 0\\
\cline{3-6}
 &  & \texttt{surgery} & -0.36 & -0.36 & 0\\
\cline{3-6}
 &  & \texttt{surgery:elix\_score} & 0 & $1 \times 10^{-1}$ & 0\\
\cline{3-6}
 &  & \texttt{bmi} & 0 & 0 & 0.10\\
\cline{3-6}
\multirow{-6}{*}{\centering\arraybackslash Outcome} & \multirow{-6}{*}{\centering\arraybackslash $\bm\omega$} & \texttt{hgba1c} & 0 & 0 & 0.18\\
\cline{1-6}
 &  & \texttt{(Intercept)} & -2.44 & -2.44 & -1.10\\
\cline{3-6}
 &  & \texttt{elix\_score} & 0.80 & 0.50 & 0.60\\
\cline{3-6}
 &  & \texttt{site} & 0.50 & 0.25 & -0.55\\
\cline{3-6}
 &  & \texttt{smoking\_status[former]} & -0.50 & -0.50 & 0\\
\cline{3-6}
 &  & \texttt{smoking\_status[current]} & -1.00 & -1.00 & 0\\
\cline{3-6}
\multirow{-6}{*}{\centering\arraybackslash Missing Eligibility} & \multirow{-6}{*}{\centering\arraybackslash $\bm\rho$} & \texttt{insulin} & 0 & 0 & 0.50\\
\hline
\end{tabular}\caption{Coefficients for simulation settings, summarized for each missing data mechanism in Figure 3. Component variables comprising $\bm L_{k0}$ include \texttt{gender}, \texttt{race}, \texttt{insulin} (usage), \texttt{smoking\_status} and \texttt{elix\_score} (comorbidities). Component variables comprising $\bm L_{kt}$ include \texttt{bmi}, \texttt{hgba1c}. Treatment variables are denoted by \texttt{surgery} ($A_{kt}$) and \texttt{bs\_type} ($A'_{kt}$)}\label{table:sim_coef}
\end{table}

\section{Data Application}

\subsection{Estimation of Component Inverse Probability Weights}
\label{sec:data_ipw}

In our data application, both intention-to-treat and per-protocol effects were estimated. It's worth pointing out that in this context, non-adherence entailed non-surgical patients in trial $m$ getting surgery at some time $t > 0$. That is $A_{mkt} = 1 \neq A_{mk} = 0$ for some $t > 0$. Non-adherence was one-sided, as one could not undo the treatment of bariatric surgery. In other contexts, adherence might refer to the recommendation to undergo surgery, not the initiation of treatment itself. While intention-to-treat effects under this setting may be of greater clinical interest than ITTs under our setting, we are unable to answer such questions due to lack of data granularity. Nevertheless, we computed both estimands for illustrative purposes.

For the purposes of eligibility ascertainment, and subsequent confounding adjustment, $\bm L^{e}_{mk}$ consisted of BMI, blood glucose measures (A1c or fasting blood glucose), and insulin usage/diabetic prescriptions. Critically, these variables could not be used to adjust for the possibility of selection bias. Additional non-eligibility defining covariates, $\bm L^c_{mk}$, included sex, health care site, age, race, estimated glucose filtration rate (eGFR), and self-reported smoking status. eGFR was estimated from serum-creatinine values via the CKD-EPI creatinine equation \citep{ckd-epi2021}. Precise lists of ICD-9 and CPT codes used to define outcomes can be found in the original study by O'Brien et al. \citep{obrien2018microvascular}. 

Each component weight model was estimated using a pooled logistic regression model with the following variables: 

\begin{itemize}
   \item \textbf{Selection bias weights} $(\bm W^R)$: health care site, baseline age, sex, smoking status, baseline eGFR, surgery
    \item \textbf{Confounding weights} $(\bm W^A)$: health care site, baseline age, sex, smoking status, baseline eGFR, baseline BMI, baseline A1c, baseline insulin usage
    \item \textbf{Non-adherence weights} $(\bm W^N)$: health care site, time-varying age, sex, time-varying smoking status, time-varying eGFR, time-varying BMI, time-varying A1c, baseline insulin usage
\end{itemize}

Intention-to-treat effects were estimated using weights for confounding and selection bias, while per-protocol effects were estimated with additional weights for non-adherence. No weights were used for censoring, which is implicitly making an assumption of non-informative censoring, as made by O'Brien et al. \citep{obrien2018microvascular}. Time-varying covariates (BMI, A1c, eGRF, age) for non-adherence weights were taken to be the most recent available values.

Weight stabilization was utilized for each component weight according to the formulae in Table 1. Additionally, each component weight was truncated at the 99\% quantile, a commonly used threshold in inverse probability weighting \citep{cole2008}.

Some subjects were missing baseline eGFR. Since eGFR was not an eligibility defining covariate but nevertheless a useful confounder/covariate for addressing possible selection bias, we imputed it via a Gamma generalized linear model. BMI measurements were not required to determine eligibility for surgical patients ($A_{mk} = 1$) given the unique parallel between study and treatment eligibility. Furthermore, no blood glucose measurement was required for patients with an active diabetes prescription at the start of trial $m$. For such subjects whose eligibility for the study was partially determined by initiation of bariatric surgery and/or presence of an active diabetic medication prescription, baseline BMI/A1c values (two important confounders) could be missing. Upon restricting to eligible subject-trials, missing BMI/A1c were imputed, again using Gamma generalized linear models. The focus of this work is not missing confounders, another tricky problem which other literature has attempted to address \citep{levis2022, evans_2020, williamson2012}. Nevertheless, this missingness is worth mentioning because it demonstrates that in real data applications, particularly those in EHR-based observational studies, missing data is likely to be pervasive at many levels of an analysis.

\subsection{Sensitivity of Variance to Eligibility Lookback Times}\label{sec:supp_fig}

\begin{figure}[H]
    \centering
    \includegraphics[width = \textwidth]{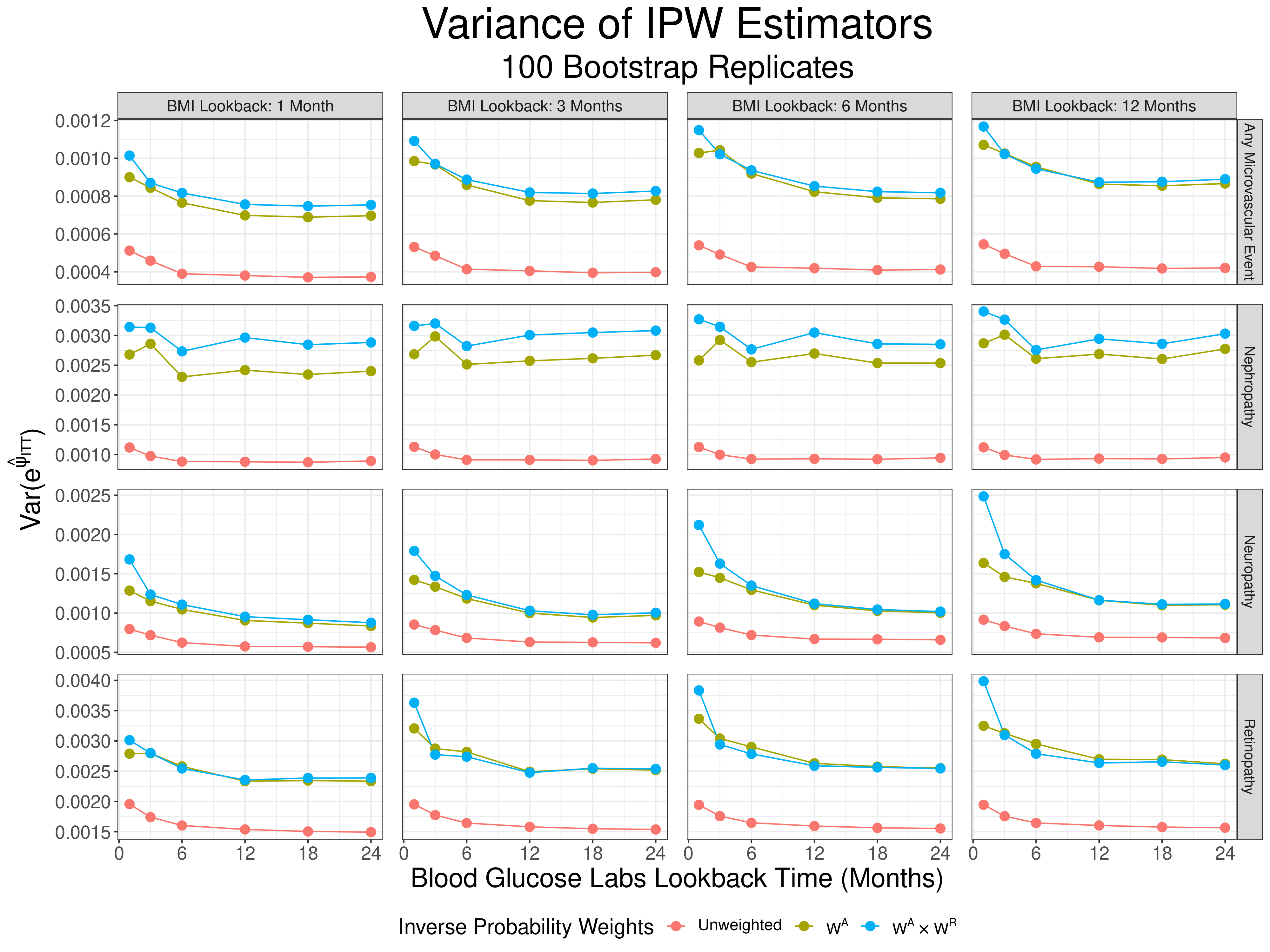}
    \caption{Variance of discrete intention-to-treat hazard ratios, $\exp(\hat\psi_{ITT})$. Variances were estimates from $B = 100$ bootstrap replicates at the subject $(k)$ level.}
    \label{fig:var}
\end{figure}

Variance estimates for discrete intention-to-treat hazard ratios, $\exp(\hat\psi_{ITT})$ over the complete grid of lookback times are plotted in Figure \ref{fig:var}. Estimates of the variance were derived from $B = 100$ bootstrap replicates at the subject $(k)$ level (see Section \ref{sec:inference} for more details). Due to the size of the data and the number of distinct lookback combinations considered, doing more than 100 replicates per panel, or bootstrapping variance for every per-protocol estimate was not computationally feasible. In general, variance tended to decrease with increasing lookback time, with plateaus mirroring those seen with sample size in Figure 4. Thus, the choice of appropriate lookback windows for determining eligibility can be viewed as a bias-variance trade off. Additionally, as seen in Supplementary Figure \ref{fig:var}, inclusion of $W^R$ for addressing selection bias did not drastically increase variance on top of weights $W^{A}$ for confounding for ITT estimates.

\section{Inference}\label{sec:inference}
Inference for the model framework outlined in this work requires care, owing to both the use of inverse probability weights and perhaps more importantly, the fact the the same subject can contribute to multiple trials, including duplicated person-time across these trials. Techniques for estimating the variance of $\hat\psi_{PP}$ (without loss of generality) in the literature can broadly be characterized into two approaches, robust/sandwich standard errors \citep{gupta2021, hernan2008, danaei2013observational} and bootstrapping \citep{dickerman2019, maringe2020reflection, hajage2022extracorporeal, emilsson2018, caniglia2019, danaei2018, dickerman2022, barda2021}.

We outline a way to obtain a closed form of the variance of $\hat\psi_{PP}$ in the Section \ref{sec:supp_var}. In practice, however, such a form would need to be re-derived based on the specific choices for component weight functions. While (pooled) logistic regressions make suitable choices for each weight model, censoring/non-adherence models may be well suited to other forms based on exponential models \citep{peskoe2021}.

As such, we recommend a bootstrapping approach, where replicates are re-sampled at the subject $(k)$ level as opposed to subject-trial or subject-trial-observation level. Though bootstrapping is common in the literature, coverage properties haven't been well studied within this framework. Section \ref{sec:sim_inference} establishes that this bootstrapping procedure for 95\% confidence intervals achieves nominal coverage.

\subsection{Validity of Bootstrap}\label{sec:sim_inference}

In order to establish the validity of re-sampling subjects in order to bootstrap variance and/or confidence intervals, we conducted an additional simulation study. For each of the three missing data scenarios in Figure 3, 1,000 datasets of size $K = 5,000$ subjects were simulated under the first hypothetical study design outline (with eligibility criteria depending on cutoffs for both BMI and A1c measurements). Note that this is exactly the same simulation set up as hypothetical study \#1 with half as many subjects $K$ to make bootstrapping a little bit more computationally feasible.

For each simulated dataset, $B = 1,000$ bootstrap replications were conducted by re-sampling entire subjects and then running the analysis procedure described in Section \ref{sec:data_ipw}. $B = 1,000$ is actually a larger number of replicates than are used in much of the literature. For example Hajage and colleagues \cite{hajage2022extracorporeal} conducted bootstrapping with $B = 200$, while Dickerman et al. \cite{dickerman2022} and Barda and coauthors \cite{barda2021} used $B = 500$, likely owing to the very large size of the analysis data set.

95\% confidence intervals for $\hat\psi_{PP}$ were estimated for each simulated dataset using three different methods: normal intervals, pivotal intervals, and percentile based intervals \cite{wasserman2004all, efron1993introduction}. As seen in Table \ref{table:boot_var}, coverage of 95\% confidence intervals roughly achieved nominal rates, confirming that bootstrapping at the subject level is an appropriate inferential procedure in this analytical framework. Due to both discretization of time and pooling of trials, data size typically limits the number of bootstrap replicates which can be conducted in sequential target trial emulations. Given that percentile based methods typically require more replicates to converge, perhaps normal or pivotal intervals are more appropriate.

\begin{table}[H]
    \centering
    \begin{tabular}{|Sc|Sc|Sc|Sc|}
       \hline
       \multicolumn{1}{|Sc|}{\multirow{2}{*}{\textbf{Setting}}} & \multicolumn{3}{Sc|}
       {\textbf{Coverage of 95\% Confidence Intervals}} \\
       \cline{2-4}
        & \textbf{Normal} & \textbf{Pivotal} & \textbf{Percentile}\\
        \hline
        M-Bias & 0.948 & 0.956 & 0.944\\
        \hline
        Treatment Effect Heterogeneity & 0.936 & 0.935 & 0.933\\
        \hline
        M-Bias w/ Mediator & 0.956 & 0.955 & 0.948\\
        \hline
        \end{tabular}
    \caption{Coverage of 95\% confidence intervals from $B = 1,000$ bootstrap replicates for $\hat\psi_{PP}$ from hypothetical study \#1. Bootstrap replicates resample subjects $(k)$ and are computed using 3 different methods.}\label{table:boot_var}
\end{table}

\subsection{Derivation of Asymptotic Variance of $\hat{\bm{\psi}}_{\textit{\textbf{PP}}}$}\label{sec:supp_var}

For ease of notation, let $\bm \psi = (\bm \psi_0, \psi_{PP})$ where $\bm \psi_0$ contains all of the parameters necessary to describe the baseline hazard in the pooled logistic regression of Equation (6). In the absence of missing data, confounding, and any sort of censoring (including due to non-adherence), one could proceed with an analysis based on solving the system of estimating equations

\begin{equation}
U^C(\bm \psi) = \frac{1}{K}\sum_{k = 1}^{K}U_k(\bm \psi) = \frac{1}{K}\sum_{k = 1}^K \bm D_k^T \bm V_k^{-1}(\bm Y_k - \bm \mu_k) = \bm 0
\end{equation}

\noindent where $\bm Y_k = (\bm Y_{1k}, \bm Y_{2k}, ..., \bm Y_{mk})$ is the collection of all outcomes a subjects contributes to all trials in the pooled dataset (with analagous notions for $\bm A_k, \bm R_k$ etc.), $\mu = \mathbb{E}[Y|A] = g_\mu^{-1}(\bm \psi_0 + A\psi_{PP})$ where $g(\cdot)_\mu$ is a user-specified link function (e.g. logit link), $\bm D_k = \partial \bm \mu_k/\partial \bm \psi$, and $\bm V_k = \text{Var}[\bm Y_k|\bm A_k]$.

Of course, due to the complex observational setting, missing eligibility data, confounding, censoring, and non-adherence all must be addressed, which we choose to do via inverse probability weighting. Assume each set of component weights in a Table 1 is modeled by 

\begin{equation}
\begin{aligned}
(W^{A}_{mk})^{-1} &= g^{-1}_{A}(\bm L_{mk}; \bm \alpha_A) \\
(W^{N}_{mkt})^{-1} &= g^{-1}_{N}(\bm L_{mkt}; \bm \alpha_N) \\
(W^{C}_{mkt})^{-1} &= g^{-1}_{C}(\bm L_{mkt}; \bm \alpha_C) \\
(W^{R}_{mk})^{-1} &= g^{-1}_{R}(\bm L^{c}_{mk}; \bm \alpha_R) \\
\end{aligned}
\end{equation}

\noindent where $g_A(\cdot), g_N(\cdot), g_C(\cdot)$ and $ g_R(\cdot)$ are user specified link functions. Commonly, these may be logit links, enabling a pooled logistic regression approach towards estimating weights, as we do in our simulations. This need not be the case however, as censoring/non-adherence weights may be more appropriately modeled via a Cox model or exponential model \citep{peskoe2021}.

Given consistent estimators for each set of coefficients for modeling weights, say 
$\hat{\bm \alpha}_A,  \hat{\bm \alpha}_N,  \hat{\bm \alpha}_C, \hat{\bm \alpha}_R$ the estimator $\hat{\bm \psi}$ is then solution to the generalized estimating equation (GEE) 

\begin{equation}
\begin{aligned}
U(\bm \psi, \hat{\bm \alpha}_A,  \hat{\bm \alpha}_N,  \hat{\bm \alpha}_C,  \hat{\bm \alpha}_R) &= \frac{1}{K}\sum_{k = 1}^{K} U_k(\bm \psi, \hat{\bm \alpha}_A,  \hat{\bm \alpha}_N,  \hat{\bm \alpha}_C,  \hat{\bm \alpha}_R)\\ 
&= \frac{1}{K}\sum_{k = 1}^{K}\bm D_k^T \bm V_k^{-1}\bm \Delta_k(\hat{\bm \alpha}_A,  \hat{\bm \alpha}_N,  \hat{\bm \alpha}_C,  \hat{\bm \alpha}_R)(\bm Y_k - \bm \mu_k) \\ &= \bm 0
\end{aligned}
\end{equation}

\noindent where $\bm \Delta_k$ is an $n_k \times n_k$ diagonal matrix with entries $R_{mk}W_{mkt}$ and $n_k$ is the total number of rows a subject contributes to the overall pooled dataset (e.g. the length of $\bm Y_k$).

Taking a Taylor series expansion around the true set of parameters $(\bm \psi^*, \bm \alpha^*_A, \bm \alpha^*_N, \bm \alpha^*_C, \bm \alpha^*_R)$ and multiplying by $\sqrt{K}$, we have under standard regularity conditions \cite{van2000asymptotic}.

\begin{equation}\label{eq:taylor_exp}
\begin{aligned}
\bm 0 &= \frac{1}{\sqrt{K}}\sum_{k = 1}^K U(\bm \psi^*, \bm \alpha^*_A, \bm \alpha^*_N, \bm \alpha^*_C, \bm \alpha^*_R) \\ 
&~~~+ \frac{\partial}{\partial \bm\psi }\mathbb{E}[U_k(\bm \psi, \bm \alpha^*_A, \bm \alpha^*_N, \bm \alpha^*_C, \bm \alpha^*_R)]\biggr|_{\bm\psi = \bm\psi^*} \sqrt{K}(\hat{\bm\psi} - \bm \psi^*) \\
&~~~+ \frac{\partial}{\partial \bm\alpha_{A} }\mathbb{E}[U_k(\bm \psi^*, \bm \alpha_A, \bm \alpha^*_N, \bm \alpha^*_C, \bm \alpha^*_R)]\biggr|_{\bm\alpha_{A} = \bm\alpha_{A}^*}\sqrt{K}(\hat{\bm\alpha}_A - {\bm\alpha_A^*}) \\
&~~~+ \frac{\partial}{\partial \bm\alpha_{N} }\mathbb{E}[U_k(\bm \psi^*, \bm \alpha_A^*, \bm \alpha_N, \bm \alpha^*_C, \bm \alpha^*_R)]\biggr|_{\bm\alpha_{N} = \bm\alpha_{N}^*}\sqrt{K}(\hat{\bm\alpha}_N - {\bm\alpha_N^*}) \\
&~~~+ \frac{\partial}{\partial \bm\alpha_{C} }\mathbb{E}[U_k(\bm \psi^*, \bm \alpha_A^*, \bm \alpha_N^*, \bm \alpha_C, \bm \alpha^*_R)]\biggr|_{\bm\alpha_{C} = \bm\alpha_{C}^*}\sqrt{K}(\hat{\bm\alpha}_C - {\bm\alpha_C^*}) \\
&~~~+ \frac{\partial}{\partial \bm\alpha_{R} }\mathbb{E}[U_k(\bm \psi^*, \bm \alpha_A^*, \bm \alpha_N^*, \bm \alpha_C^*, \bm \alpha_R)]\biggr|_{\bm\alpha_{R} = \bm\alpha_{R}^*}\sqrt{K}(\hat{\bm\alpha}_R - {\bm\alpha_R^*}) \\
&~~~+o_P(1)
\end{aligned}
\end{equation}

Let $S_k^A(\bm\alpha_A^*)$ be the contribution of the $k^{th}$ individual to the score for the model used to estimate $\bm\alpha_A$, with analogous notions for each set of parameters used in the weight models. We perform similar Taylor series expansions of the estimating equations used to obtain each set of $\bm\alpha$ parameters and upon rearranging terms, are left with the following expressions.

\begin{equation}\label{eq:weight_scores}
\begin{aligned}
\sqrt{K}(\hat{\bm\alpha}_A - {\bm\alpha_A^*}) &= -\frac{1}{\sqrt{K}}\sum_{k = 1}^K \biggr[\frac{\partial}{\partial\bm \alpha_A} \mathbb{E}[S_k^{A}(\bm \alpha_A)] \biggr |_{\bm\alpha_A = \bm \alpha_A^*} \biggr]^{-1}\bm R_k^T S_k^A(\bm \alpha_A^*)+o_P(1) \\
\sqrt{K}(\hat{\bm\alpha}_N - {\bm\alpha_N^*}) &= -\frac{1}{\sqrt{K}}\sum_{k = 1}^K \biggr[\frac{\partial}{\partial\bm \alpha_N} \mathbb{E}[S_k^{N}(\bm \alpha_N)] \biggr |_{\bm\alpha_N = \bm \alpha_N^*} \biggr]^{-1}\bm R_k^T S_k^N(\bm \alpha_N^*)+o_P(1) \\
\sqrt{K}(\hat{\bm\alpha}_C - {\bm\alpha_C^*}) &= -\frac{1}{\sqrt{K}}\sum_{k = 1}^K \biggr[\frac{\partial}{\partial\bm \alpha_C} \mathbb{E}[S_k^{C}(\bm \alpha_C)] \biggr |_{\bm\alpha_C = \bm \alpha_C^*} \biggr]^{-1}\bm R_k^T S_k^C(\bm \alpha_C^*) +o_P(1)\\
\sqrt{K}(\hat{\bm\alpha}_R - {\bm\alpha_R^*}) &= -\frac{1}{\sqrt{K}}\sum_{k = 1}^K \biggr[\frac{\partial}{\partial\bm \alpha_R} \mathbb{E}[S_k^{R}(\bm \alpha_R)] \biggr |_{\bm\alpha_R = \bm \alpha_R^*} \biggr]^{-1} S_k^R(\bm \alpha_R^*) +o_P(1)\\
\end{aligned}
\end{equation}

Substituting Equation \ref{eq:weight_scores} into Equation \ref{eq:taylor_exp} and rearranging yields

\begin{equation}
\begin{aligned}
 \sqrt{K}(\hat{\bm\psi} - \bm \psi^*)  = -J^{-1}\frac{1}{\sqrt{K}}\biggr[&\sum_{k = 1}^K U(\bm \psi^*, \bm \alpha^*_A, \bm \alpha^*_N, \bm \alpha^*_C, \bm \alpha^*_R) \\ &- Q_AI_A^{-1}\sum_{k = 1}^K \bm R_k^T S_k^A(\bm \alpha_A^*) \\
 &- Q_N I_N^{-1}\sum_{k = 1}^K \bm R_k^T S_k^N(\bm \alpha_N^*) \\
 &- Q_C I_C^{-1}\sum_{k = 1}^K \bm R_k^T S_k^C(\bm \alpha_C^*) \\
 &- Q_R I_R^{-1}\sum_{k = 1}^K S_k^R(\bm \alpha_R^*) 
 \biggr]   +o_P(1)
\end{aligned}
\end{equation}

\noindent where

\begin{equation}
\begin{aligned}
J &= \frac{\partial}{\partial \bm\psi }\mathbb{E}[U_k(\bm \psi, \bm \alpha^*_A, \bm \alpha^*_N, \bm \alpha^*_C, \bm \alpha^*_R)]\biggr|_{\bm\psi = \bm\psi^*} \\
Q_A &= \frac{\partial}{\partial \bm\alpha_{A} }\mathbb{E}[U_k(\bm \psi^*, \bm \alpha_A, \bm \alpha^*_N, \bm \alpha^*_C, \bm \alpha^*_R)]\biggr|_{\bm\alpha_{A} = \bm\alpha_{A}^*}\\
Q_N &= \frac{\partial}{\partial \bm\alpha_{N} }\mathbb{E}[U_k(\bm \psi^*, \bm \alpha_A^*, \bm \alpha_N, \bm \alpha^*_C, \bm \alpha^*_R)]\biggr|_{\bm\alpha_{N} = \bm\alpha_{N}^*} \\
Q_C &=\frac{\partial}{\partial \bm\alpha_{C} }\mathbb{E}[U_k(\bm \psi^*, \bm \alpha_A^*, \bm \alpha_N^*, \bm \alpha_C, \bm \alpha^*_R)]\biggr|_{\bm\alpha_{C} = \bm\alpha_{C}^*} \\
Q_R &= \frac{\partial}{\partial \bm\alpha_{R} }\mathbb{E}[U_k(\bm \psi^*, \bm \alpha_A^*, \bm \alpha_N^*, \bm \alpha_C^*, \bm \alpha_R)]\biggr|_{\bm\alpha_{R} = \bm\alpha_{R}^*}  \\
I_A &= \frac{\partial}{\partial\bm \alpha_A} \mathbb{E}[S_k^{A}(\bm \alpha_A)] \biggr |_{\bm\alpha_A = \bm \alpha_A^*}  \\
I_N &= \frac{\partial}{\partial\bm \alpha_N} \mathbb{E}[S_k^{N}(\bm \alpha_N)] \biggr |_{\bm\alpha_N = \bm \alpha_N^*} \\
I_C &= \frac{\partial}{\partial\bm \alpha_C} \mathbb{E}[S_k^{C}(\bm \alpha_C)] \biggr |_{\bm\alpha_C = \bm \alpha_C^*}\\
I_R &= \frac{\partial}{\partial\bm \alpha_R} \mathbb{E}[S_k^{R}(\bm \alpha_R)] \biggr |_{\bm\alpha_R = \bm \alpha_R^*} 
\end{aligned}
\end{equation}

Then, by the a Central Limit theorem argument, we have that 
\begin{equation}
\begin{aligned}
\sqrt{K}\bigr(\hat{\bm\psi} - \bm\psi^*\bigr) &\to \text{Normal}(0, \Omega)\\
\Omega &= J^{-1}\Gamma J^{-1} \\
\Gamma &= \text{Var}\biggr[U_k(\bm \psi, \bm \alpha^*_A, \bm \alpha^*_N, \bm \alpha^*_C, \bm \alpha^*_R)  - Q_A I_A^{-1} S^A(\bm \alpha_A^*)  
\\ 
&~~~~~~~~~- Q_N I_N^{-1} S^N(\bm \alpha_N^*)  - Q_C I_C^{-1} S^C(\bm \alpha_C^*) - Q_R I_R^{-1} S^R(\bm \alpha_R^*)    \biggr]
\end{aligned}
\end{equation}

\clearpage
\bibliographystyle{ama}
\bibliography{ref}